\documentclass[aps,
reprint,
superscriptaddress
]{revtex4-2}
\usepackage{amsmath,amssymb,physics,bm}
\usepackage{graphicx, microtype,siunitx} 
\usepackage[colorlinks=true,linkcolor=blue,citecolor=blue,urlcolor=blue]{hyperref}
\usepackage[all]{hypcap} 
\usepackage[colorinlistoftodos]{todonotes}

\usepackage{mathptmx}

\newcommand{\vvec}[0]{{\bm v}}
\newcommand{\xivec}[0]{{\bm \xi}}
\newcommand{\nvec}[0]{{\bm n}}
\newcommand{\Evec}[0]{{\bm E}}
\newcommand{\vx}[0]{{\it v}}
\newcommand{\xx}[0]{{\it x}} 

\definecolor{cream}{RGB}{222,217,201}

\begin{document}

\title[]{Particle-based model of active skyrmions}

\author{Allison W. Teixeira}
\affiliation{Centro de Física Teórica e Computacional, Faculdade de Ciências, Universidade de Lisboa, 1749-016 Lisboa, Portugal.}
 \affiliation{Departamento de Física, Faculdade de Ciências, Universidade de Lisboa, P-1749-016 Lisboa, Portugal.}
\author{Mykola Tasinkevych}
\email{mykola.tasinkevych@ntu.ac.uk}
  \affiliation{Centro de Física Teórica e Computacional, Faculdade de Ciências, Universidade de Lisboa, 1749-016 Lisboa, Portugal.}
 \affiliation{Departamento de Física, Faculdade de Ciências,
Universidade de Lisboa, P-1749-016 Lisboa, Portugal.}
\affiliation{SOFT Group, School of Science and Technology, Nottingham Trent University, Clifton Lane, Nottingham NG11~8NS, United Kingdom.}
\affiliation{International Institute for Sustainability with Knotted Chiral Meta Matter, Hiroshima University, Higashihiroshima 739-8511, Japan.}
\author{Cristóvão S. Dias}
\affiliation{Centro de Física Teórica e Computacional, Faculdade de Ciências, Universidade de Lisboa, 1749-016 Lisboa, Portugal.}
 \affiliation{Departamento de Física, Faculdade de Ciências, Universidade de Lisboa, P-1749-016 Lisboa, Portugal.}

\date{\today}

\begin{abstract}
\noindent{\bf \large Abstract} 

\noindent  Motivated by recent experimental results that reveal rich collective dynamics of thousands-to-millions of active liquid crystal skyrmions we have developed a coarse grained particle-based model of the dynamics of skyrmions in dilute regime.
The basic physical mechanism of the skyrmion motion is related to the non-reciprocal rotational dynamics of the liquid crystal director field when the electric field is turned {\it on} and {\it off}. Guided by fine grained results of the Frank-Oseen continuum approach, we have mapped this non-reciprocal director distortions onto an effective force acting asymmetrically upon switching the electrical field {\it on} or {\it off}. The coarse grained model correctly reproduces the skyrmion dynamics, including the velocity reversal as a function of the frequency of a pulse width modulated driving voltage. We have also obtained approximate analytical expressions for the phenomenological model parameters encoding their dependence upon the cholesteric pitch and the strength of the electric field. This has been achieved by fitting coarse grained skyrmion trajectories to those determined in the framework of the Frank-Oseen model. 
\end{abstract}

\maketitle


\noindent {\large \bf Introduction}

\noindent Chiral nematic liquid crystals (LCs) are known to host stable solitonic configurations \cite{Smalyukh2010,Ackerman2014,Ackerman2017,Sohn2018,Sohn2019a,Sohn2019,Song2021,Bogdanov2003,Duzgun2018,Duzgun2021,Duzgun2022,Coelho2022,Long2021} dubbed skyrmions, which are low-dimensional analogs of Skyrme solitons in nuclear physics \cite{Skyrme1962}.
Skyrmions are spatially localised topological configurations of the liquid crystal director field which exhibit particle-like behaviour, such as effective interactions between skyrmions \cite{Sohn2019a,Sohn2019} and with colloidal particles \cite{Sohn2018}. Similar structures in magnetic systems have been studied intensively over several decades 
\cite{Muhlbauer:2009,Yu2010,Zhang2016,Liu:2018,Sutcliffe_2018,Das2019,PhysRevE.90.042502,PhysRevE.97.062706,Foster2019,Teixeira_2021}. Magnetic skyrmions are envisioned as a physical realisation of bits in future memory storage and logic devices \cite{Krause2016}.

In confined geometries, LC skyrmions undergo translational motion when subject to time dependent electric fields \cite{Ackerman2017,Sohn2018}. The resulting velocity can be controlled by the frequency, strength and the duty cycle of the pulse width modulated electric field. Driven LC skyrmions represent a new class of active particles whose distinctive feature is the absence of net mass transport \cite{Ackerman2017,Sohn2018}. The translation of these topological structures is realised by controlled reconfiguration of the LC director field in an experimental setup which is identical to that used in liquid crystal display technologies \cite{Shen2020}. This provides an opportunity to exploit the controlled motion of skyrmions for development of advanced electro-optic responsive materials \cite{Wu2022}.

Surprising emergent collective behaviour has been reported experimentally including light controlled skyrmion interactions and self-assembly \cite{Sohn2019a}, reconfigurable cluster formation and formation of large-scale skyrmion crystals mediated by out-of-equilibrium elastic interactions \cite{Sohn2019}. At high skyrmion packing fractions, hexagonal crystallites of skyrmions revealed coherent motion along a certain direction resulting in an increased hexatic order parameter \cite{Sohn2020}. Experiments also show that active skyrmions can entrap and transport colloidal particles \cite{Sohn2018}. This may be used for controlled non-contact manipulation of colloids, enabling the development of advanced applications.

Despite the extensive body of experimental research, understanding the many-body dynamics of LC skyrmions remains poorly understood. Existing numerical investigations are based on minimisation of the Frank-Oseen \cite{Ackerman2017,Sohn2018} and Landau-de Gennes \cite{Duzgun2018,Duzgun2021,Duzgun2022} free energies have successfully reproduced several experimental results regarding the structure and dynamics of skyrmionic excitation. However, these fine grained approaches are limited to a small number of skyrmions, and to study emergent non-equilibrium behaviour of a large number of skyrmions new coarse graining strategies must be developed. 

The present work introduces a coarse grained particle-based model of the dynamics of an isolated skyrmion, i.e. neglecting skyrmion interactions. The model is based on time dependent forces which encode the non-reciprocal responses of the LC director field to switching the electric field {\it on} and {\it off}. The functional form of the forces are derived based on the results of the fine grained Frank-Oseen model of the skyrmion dynamics. The particle-based model developed here correctly captures the main phenomenology, including the velocity reversal as a function of the frequency of a pulse width modulated external electric field. The results on the average skyrmion velocity discussed here are applicable for long enough duration of the {\it on} and {\it off} states of the electric field. These limitations are not conceptual and can easily be overcome, as is discussed in the final section.   
\vspace{0.5cm}

\noindent {\large \bf Results}

\noindent{\bf Fine grained description of the skyrmion dynamics.} In experiments, the skyrmions are stabilized in thin films of a cholesteric LC placed between two parallel plates with homeotropic anchoring conditions. The plate separation is of the order of the cholesteric pitch $P$, which results in an unwound uniform configuration of the LC director, aligned perpendicular to the plates. Under certain conditions, this stressed configuration may be relaxed by incorporating solitonic configurations, e.g. skyrmions \cite{Ackerman2017}. These localised configurations are three dimensional topological solitons where the skyrmion tube is introduced between the two plates along the perpendicular axis, terminated by two hedgehog defects \cite{Smalyukh2010}. Applying step-like time dependent voltage across the cell results in an effective displacement of the skyrmions in a random direction, and the speed and the direction of the motion are controlled by the temporal characteristics of the electric field \cite{Ackerman2017}.  

To get insight into the dynamics of an isolated skyrmion, we resort to the underlying continuum model of liquid crystals based on the Frank-Oseen formalism. The free energy minimizing director configurations are obtained numerically by integrating the corresponding evolution equation from the LC director field (see Methods for details). 

\begin{figure}[t!]
\centering
\includegraphics[width=0.95\columnwidth]{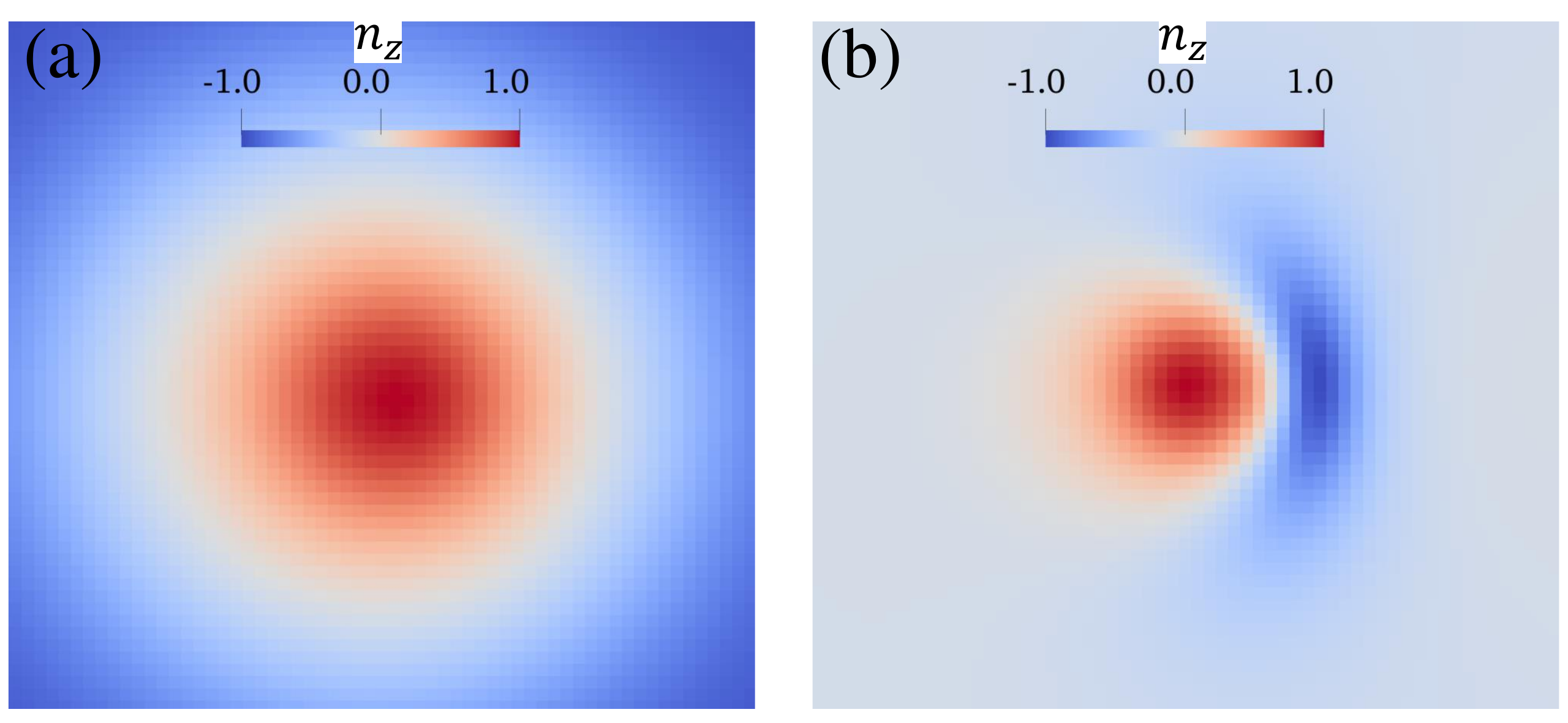} 
\caption{{\bf Equilibrium director configurations}. colour coded cross-sectional representation of the $z-$component $n_z$ of the equilibrium  nematic director, shown in the middle plane $z = L_z/2$, where $L_z$ is the distance between the confining plates. (a) Axially symmetric configuration obtained at zero external electric field. This 2D configuration has the structure of a skyrmion, while the full $3D$ configuration corresponds to that of a so-called toron \cite{Smalyukh2010}, here we do not use this terminology. (b) Bimeron-like cross-sectional configuration obtained at moderate electric field corresponding to the electric potential drop of $3.5 V$. The electric field is applied perpendicular to the plane of cross sections.
}
\label{fig_configs_FO_model}
\end{figure} 

\begin{figure}[t!]
\centering
\includegraphics[width=0.95\columnwidth]{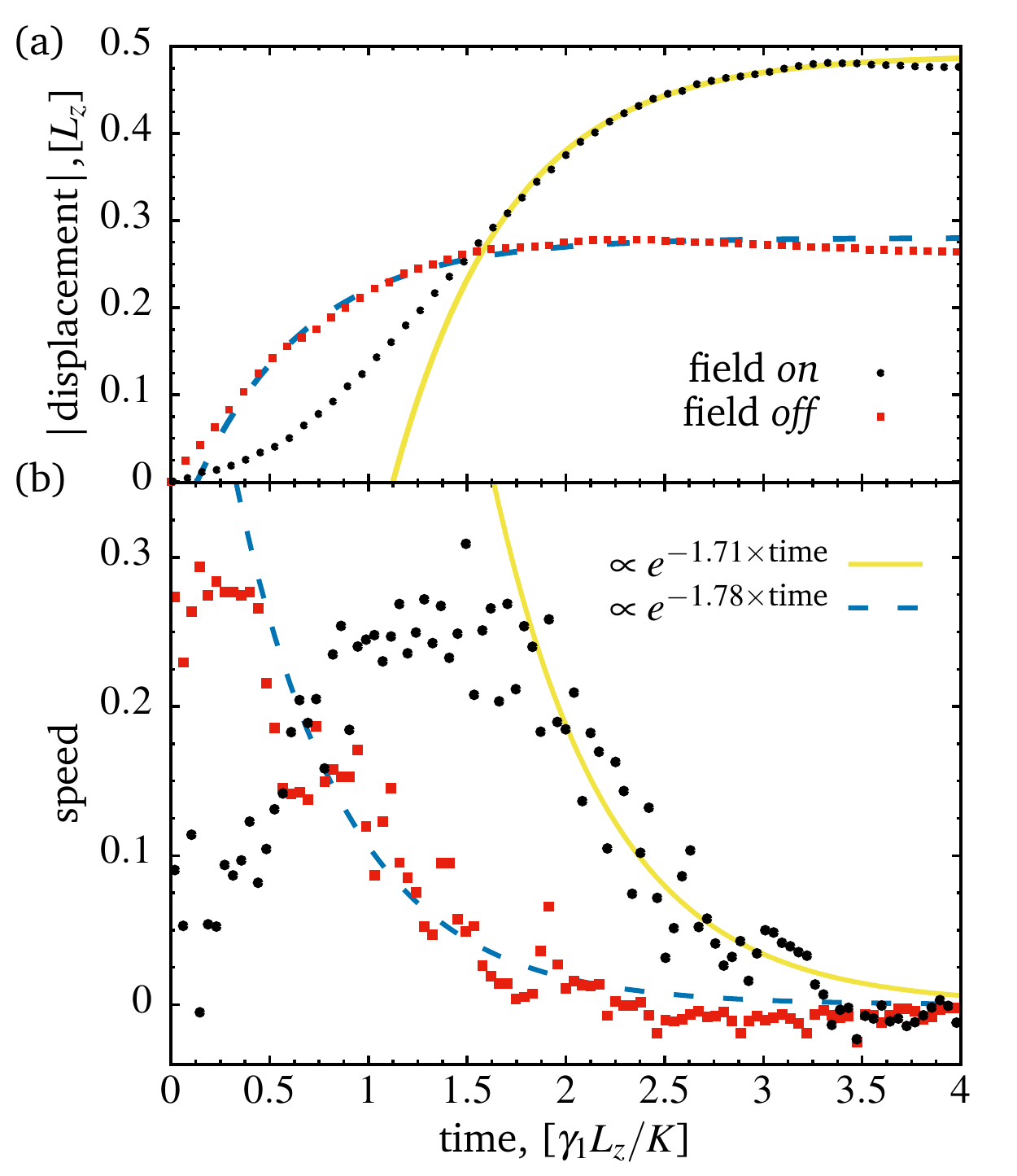} 
  \caption{{\bf Electric field driven skyrmion motion.} (a) Absolute value of the relative displacement of the $x-$component of the skyrmion's ``center of mass'', defined in Eq.~(\ref{toron-center-of-mass}), as a function of time. The red solid squares correspond to the field {\it off} state when the axially symmetric skyrmion configuration morphs into the bimeron-like configuration. The black solid circles represent the relative skyrmion displacement during the field {\it on} state. The starting configuration was the stable bimeron one obtained in the previous field {\it on} state. The dashed blue and solid yellow curves are exponential fits to the numerical data. (b) The skyrmion's speed obtained by numerical differentiation of the corresponding curves in (a). The colour code matches that in (a).   
  }
  \label{Fig2}
\end{figure}

 Fig.~\ref{fig_configs_FO_model} demonstrates equilibrium director configurations corresponding to field {\it off}, Fig.~\ref{fig_configs_FO_model}(a), and field {\it on}, Fig.~\ref{fig_configs_FO_model}(b) states. The later structure is equivalent to so-called "bimeron" configuration known in magnetic skyrmion literature \cite{Li2020}. In a course of the free energy minimization, we compute skyrmion displacement $x_c(t)-X_0$ as a function of time $t$, where the skyrmion's center of mass $x_c$ is defined in the Methods section. A typical trajectory along the field {\it on} state is shown in Fig.~\ref{Fig2}(a) by solid black circles. In this case, the skyrmion moves in the positive $x-$direction, and after fast initial stage, the skyrmion approaches the stationary state exponentially (see solid yellow fitting curve on Fig.~\ref{Fig2}(a)). After the equilibrium is reached, the electric field is switched off and the bimeron configuration morphs back to the axisymmetric one. During this process, the $\mathbb{S}^2$ north pole preimage retracts, i.e. moves in the negative $x-$direction. The corresponding displacement is depicted in Fig.~\ref{Fig2}(a) by the red solid squares, and the dashed blue curve represents an exponential fit to the late time dynamics, shown in Fig.~\ref{fig_configs_FO_model}(a). The net displacement during the {\it off} state is roughly twice as large as during the {\it on} state. Fig.~\ref{Fig2}(b) shows the skyrmion speed as a function of time along the {\it on} and {\it off} states. We assume that the skyrmion velocity is positive/negative for the {\it on}/{\it  off} state, respectively.

\vspace{0.5cm}

\noindent {\bf Coarse grained skyrmion's dynamics.} In a coarse grained model, we characterise the skyrmion by its center of mass coordinates. The center of mass undergoes translational motion in response to turning the electric field {\it on} and {\it off}, with the velocity $\vvec'(t')$, which depends exponentially on time $t'$. To account for this effect we introduce an effective time-dependent force $F'_{n}(t') \hat{\vvec}(t')$, which is proportional to the unit vector $\hat{\vvec}(t')$ in the velocity direction. Based on the results of the previous section we postulate: 
\begin{align}
F'_{n}(t') = A'_{n} e^{-\alpha'_{n} t'} = 
\left\{
\begin{array}{c}
F'_{on}(t') = A'_{on} e^{-\alpha'_{on} t'}\\
F'_{off}(t') = A'_{off} e^{-\alpha'_{off} t'}
\end{array}
\right. ,
\label{force}
\end{align}
where $n \in \{on,off\}$, and distinguishes the case when the electric field is $on$ from the case when the electric field is $off$. We find convenient to introduce ratios of the force amplitudes $A_{r}=A'_{off}/A'_{on}$; $A_{r}<0$, and the exponents $c_{r}=\alpha'_{off}/ \alpha'_{on}$; also recall that $A'_{on}>0$. In an overdamped regime this force will result in particle's velocity $\vvec'(t') = F'_{n}(t') \hat{\vvec}(t') / \gamma$, where $\gamma$ is the friction parameter. With this, the particle position is then obtained from the following system of equations:
\begin{equation}
 m \dot{\vvec'}(t') = 
- \gamma \vvec'(t')
+ F'_{n}(t') \hat{\vvec}(t') + \xivec'(t',\cal{T}'),
\label{equation_of_motion}
\end{equation}
where $m$ is an effective skyrmion mass, and for generality we also introduce random force 
$\xivec'(t',\cal{T}')$ to account for effects of thermal fluctuations, with $\cal{T}'$ denoting the temperature. We emphasize that Eq.~(\ref{equation_of_motion}) is expected to be a good approximation for low frequencies of the pulse width modulated electric field. 


We define reduced quantities as $t=t'\gamma/m$, $\vvec = \vvec' \gamma / A'_{on}$, $F_{n}=F'_{n}/A'_{on}$, $\xivec(t,{\cal T}')=\xivec'(t,{\cal T}')/A'_{on}$, $\alpha_{n}=\alpha'_{n} m/\gamma$, and rewrite equation (\ref{equation_of_motion}) in a dimensionless form

\begin{align}
&
\dot{\vvec}(t) = 
- \vvec(t)
+ F_{n} ( t ) \hat{\vvec}(t) + \xivec(t,\cal{T}'),
\label{eq_of_motion} \\
&
F_{on}(t) = e^{-\alpha_{on} t},
\nonumber \\
&
F_{off}(t) = A_r e^{-\alpha_{off} t},
\end{align}

\begin{figure}[t!]
\centering
\includegraphics[width=0.95\columnwidth]{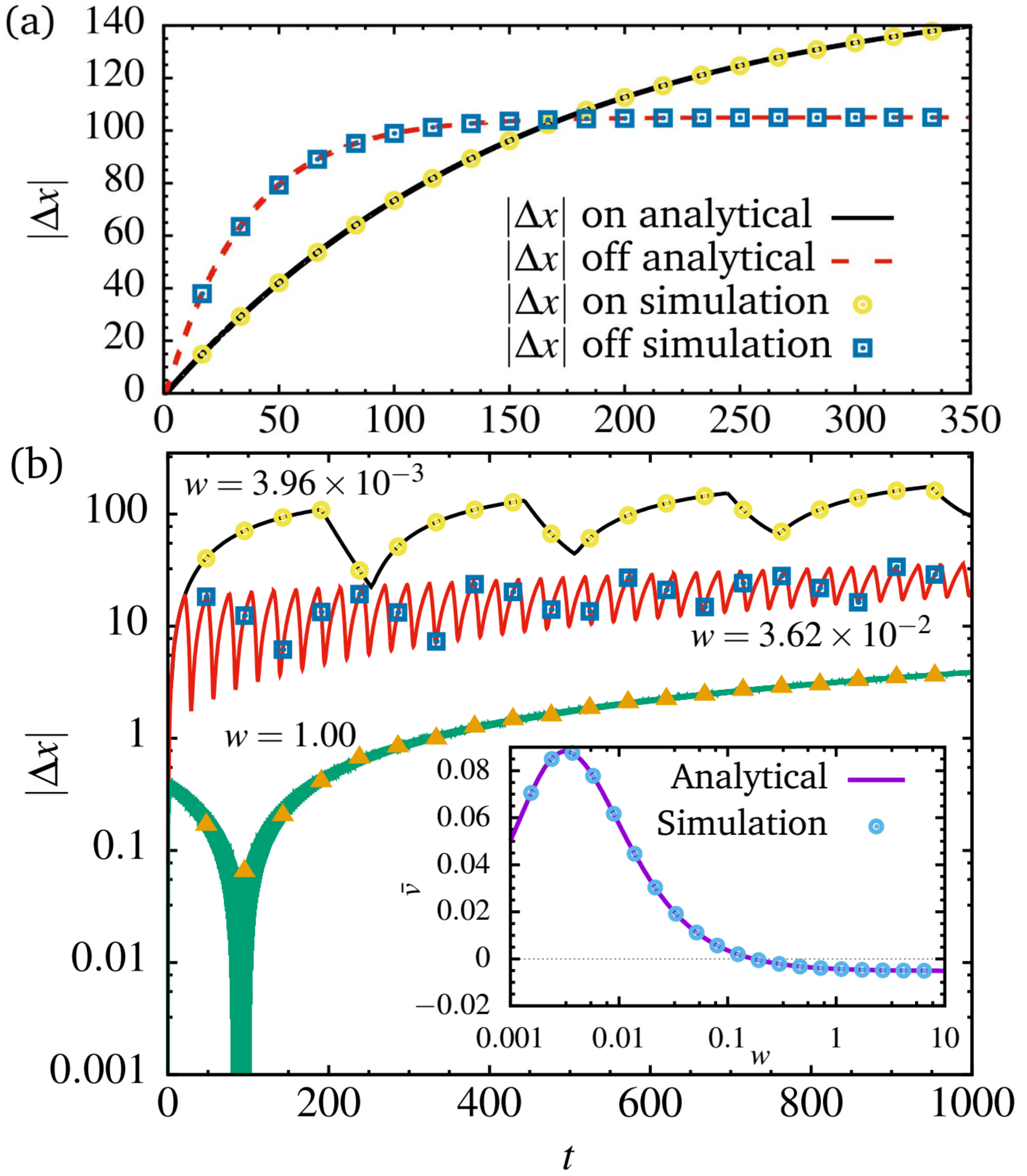} 
  \caption{{\bf Analytical results for the skyrmion's motion.} (a) Analytical (lines) and numerical (symbols) results for the absolute displacement of the skyrmion during states when the field is {\it on} (black solid line and orange circle) and {\it off} (red dashed line and blue squares), as a function of time. The remaining free model parameters are adjusted by comparing the theoretical prediction (\ref{averagevelocity}) with the experimental results \citet{Ackerman2017}, as detailed in the main text, which gives $A_{r} = -3.02$, $c_{r} = 4.50$ and $\alpha_{on}=6.4 \times 10^{-3}$. (b) Absolute displacement of the skyrmion driven by a pulse width modulated electric field with the modulation frequency of $w=3.96\times 10^{-3}$, $w=3.62\times 10^{-2}$ and $w=1$ from top to bottom, respectively. Inset shows the steady state average skyrmion velocity as a function of frequency $w$, after sufficiently large number $N$ of the field transition between the {\it on} and {\it off} states. The field duty cycle $\tau=0.75$.}
  \label{Fig3}
\end{figure}

\vspace{0.5cm}

\noindent{\bf Analytical results for the skyrmion's motion.} Without loss of generality, we can define the $x-$axis in the direction of motion, reducing the problem to one dimension. In our initial analysis, we neglect thermal fluctuations by setting $\xivec (t,{\cal{T}'}) = 0$, and thus the skyrmion displacement $\Delta x_n(t_0;t) \equiv x_n(t) -x_n(t_0)$ and velocity $v_n(t)$ at time $t$ are given by the following expressions

\begin{align}
&
\hspace*{-0.7cm}
\Delta \xx_{n}(t_0;t) =
\vx_{n}(t_0) \left(
1 - e^{ t_{0} - t}
\right)
\nonumber \\
&
\hspace*{-0.7cm}
+ 
 \frac{A_{r,n}}{\left(
1 - \alpha_{n}
\right)}
\left(
e^{\left(1 - \alpha_{n} \right) t_{0} - 
t} - e^{- \alpha_{n} t_{0}} +
\frac{e^{- \alpha_{n} t_{0}} - e^{- \alpha_{n} t}}{\alpha_{n}}
\right)
\label{displacement} \\
\nonumber \\
&
\hspace*{-0.7cm}
\vx_{n}(t) = 
\vx_{n}(t_0)~e^{t_{0} - t} 
\nonumber \\
&
\hspace*{-0.7cm}
+ \frac{A_{r,n}}{\left(
1 - \alpha_{n}
\right)}
\left(
e^{- \alpha_{n} t} - 
e^{\left(1 - \alpha_{n} \right) t_{0} - t}
\right),
\label{velocity}
\end{align}
where the skyrmion coordinates are expressed in units of $A'_{on}m/\gamma^2$, $t_0$ is the initial time, $A_{r,n} = 1$ if $n=${\it on} and $A_{r,n} = A_{r}$ if $n=${\it off}.

Since the response of the skyrmion to electric switching is asymmetric, applying a pulse width modulated electric field with the period $T$ and the duty cycle $\tau$ will result in a net displacement of the skyrmion. The electric field is {\it on} for a duration $t_{on}=\tau T$ and {\it off} for duration $t_{off} = \left(1-\tau \right)T$. We defined the effective skyrmion displacement $\Delta x_{eff}(NT)$ after $N$ switches of the electric field between {\it on} and {\it off} states as follows
\begin{align}
\label{eq:delta_xeff}
\hspace*{-0.6cm}
\Delta x_{eff}(NT) =
\sum_{k=1}^{2N} \Delta x_{k}(0;t_{k});
~ v_{k}(0) = v_{k-1}(t_{k-1})
\end{align}
where $k = on$ if $k$ is odd and $k = off $ if $k$ is even, similar rule applies for the subscript $k-1$ in the last equation defining the initial velocity. Equation~(\ref{eq:delta_xeff}) assumes that every time the field is turned {\it on} or {\it off}, the force $F_{k}$ is restarted at $t_{0} = 0$. This implicitly assumes that within each of the {\it on} or {\it off} state the skyrmion reaches the equilibrium.  This assumption is valid for large $t_{on}$ and $t_{off}$. 
The sum in (\ref{eq:delta_xeff}) can be carried out which yields
\begin{align}
&\Delta x_{eff}(NT) =
C_{on}
\left[ 
D_{on} \left( 
N e^{- t_{off}} - e^{t_{on}} E_{N}
\right)
\right.
\nonumber \\
&
\left.
+ D_{off} \left( N - E_{N} \right)
\right]
+ C_{off} \left[ 
D_{on} \left( 
N - e^{T} E_{N}
\right)
\right.
\nonumber \\
&
\left.
+ 
D_{off} \left( 
N e^{-t_{on}}- e^{t_{off}} E_{N}
\right)
\right]
+ N \left( B_{on} + B_{off} \right);
\end{align}
\begin{align}
&
B_{n} = 
\frac{A_{r,n}}{1 - \alpha_{n}}
\left[ 
e^{-t_{n}} - 1 
+ \frac{1}{\alpha_{n}} \left(
1 - e^{-\alpha_{n} t_{n}}
\right)
\right],
\nonumber \\
&
C_{n} = 
\frac{A_{r,n}}{1 - \alpha_{n}} \left(
e^{-\alpha_{n} t_{n}} - e^{-t_{n}}
\right), D_{n} = \frac{1 - e^{- t_{n}}}
{1 - e^{-\left( T \right)}},
\nonumber \\
&
E_{N} = \frac{e^{-NT }-1}{1-e^{T }}.
\label{effdisplacement}
\end{align}

The effective displacement becomes linear as a function of $N$ for large $N$, such that there exists a well defined average late time velocity of the skyrmion $\bar{v}(w)=\lim_{N \rightarrow \infty} w \left[ \Delta x_{eff} (N+1) - \Delta x_{eff} (N) \right]$, where $w = 1/T$ is the field frequency. $\bar{v}(w)$ represents the average slope of the skyrmion trajectory over one period of the electric field. We obtain by using Eq.~(\ref{effdisplacement}) a closed form expression

\begin{align}
   \bar{v} = \frac{w \left[ 
   A_{r} \left(
   1 - e^{\frac{(\tau-1) \alpha_{on} c_{r}}{w}}
   \right) + c_{r} \left(
   1 - e^{-\frac{\tau \alpha_{on}}{w}}
   \right)
   \right]}{\alpha_{on} c_{r}}.
   \label{averagevelocity}
\end{align}

\begin{figure}[t!]
\centering
\includegraphics[width=0.95\columnwidth]{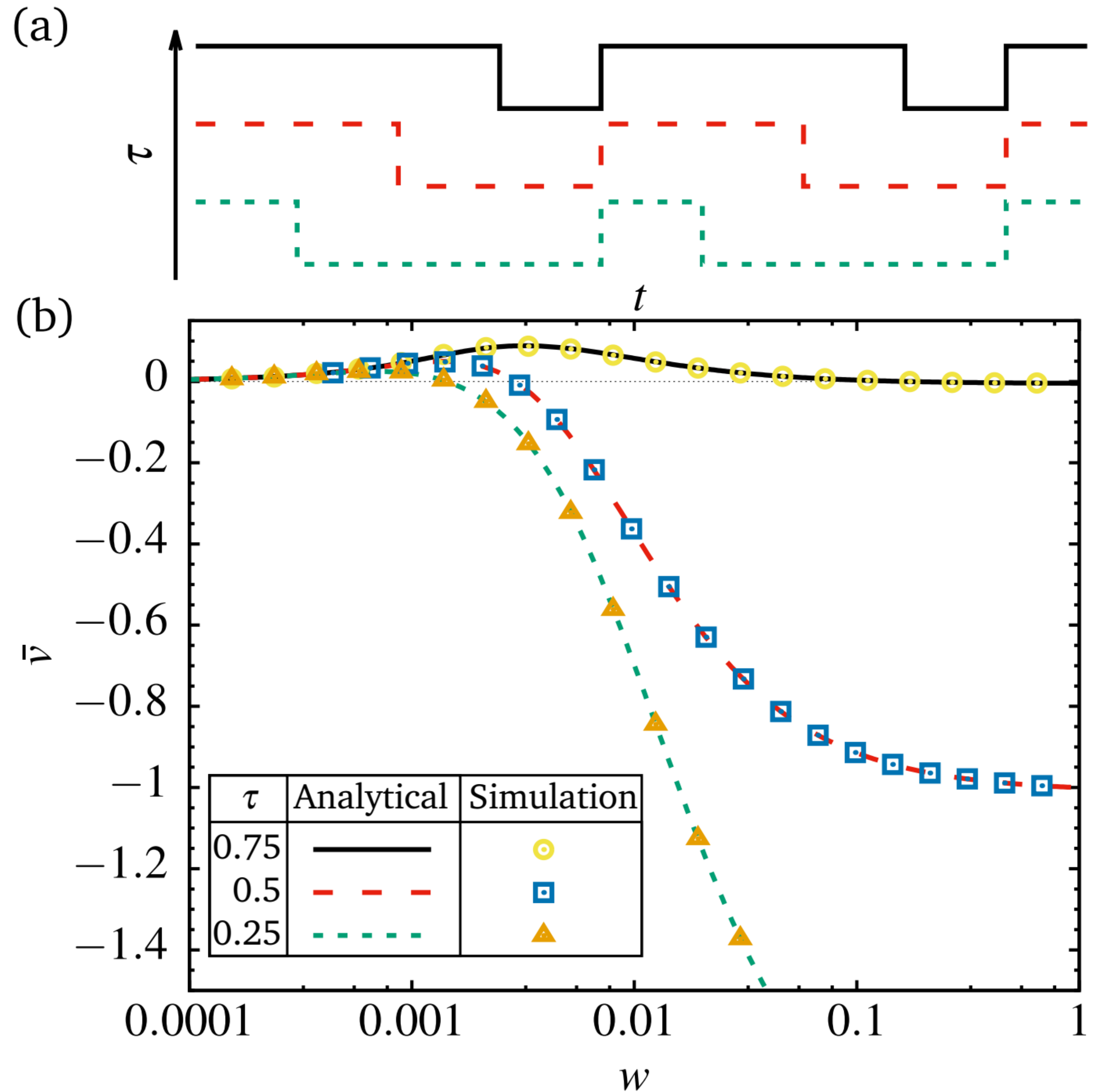} 
  \caption{{\bf Effect of the duty cycle on the skyrmion's velocity.} (a) The schematic represents the pulse width modulated electric field as a function of time $t$ for different duty cycles $\tau=0.75$, $0.5$, and $0.2$ from top to bottom, respectively. (b) Average velocity $\bar{v}$ as a function of $w$, where lines correspond to Eq.~(\ref{averagevelocity}) and symbols are results of numerical solution of Eq.~(\ref{eq_of_motion}). The remaining free model parameters are adjusted by comparing the theoretical prediction (\ref{averagevelocity}) with the experimental results \cite{Ackerman2017}, as detailed in the main text, which gives $A_{r} = -3.02$, $c_{r} = 4.50$ and $\alpha_{on}=6.4 \times 10^{-3}$. 
  }
  \label{Fig4}
\end{figure}

\begin{figure*}[t!]
\centering
\includegraphics[width=2.0\columnwidth]{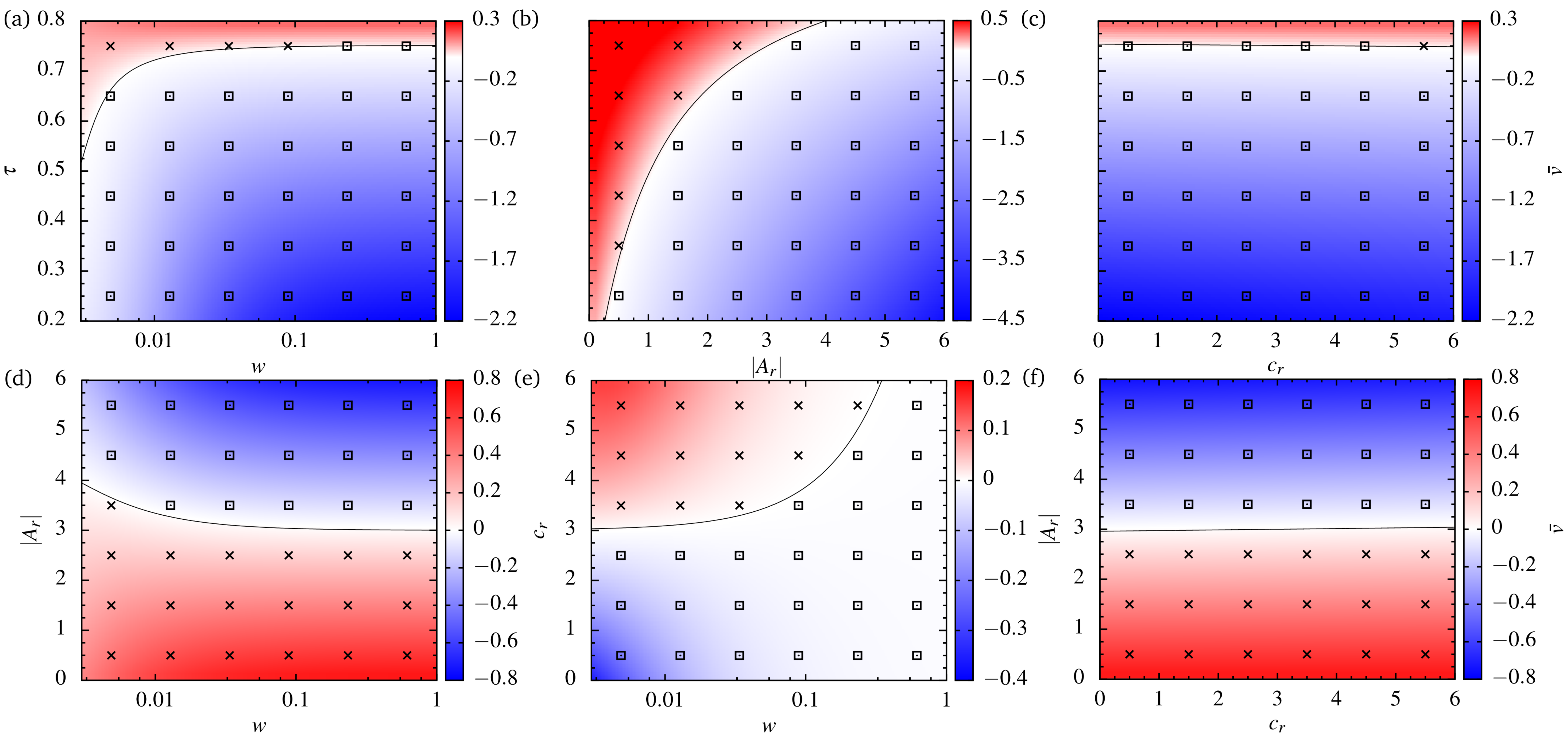} 
  \caption{{\bf Heat maps of the skyrmion's velocity.} Two-dimensional colour coded representation of the average velocity $\bar{v}$ as a function of (a) $(\tau, w)$, (b) $(\tau, |A_{r}|)$, (c) $(\tau, c_{r})$, (d) $(|A_{r}|,w)$, (e) $(c_r,w)$ and (f) $(|A_{r}|,c_r)$. The values of the parameters, if not varied along the axis of the colour maps, are set as follows  $A_{r} = -3.02$, $c_{r} = 4.50$, $\alpha_{on}=6.4 \times 10^{-3}$, $\tau = 0.75$, $w=0.173$.
The square and $\times$ symbols on the maps indicate negative and positive values of the velocity, respectively, and solid lines correspond to $\bar{v}=0$.
   }
  \label{Fig5}
\end{figure*}

The obtained equations are controlled by the three dimensionless parameters: $A_{r}$, $c_{r}$, and $\alpha_{on}$. To parameterize them, we map $\bar{v}(w)$ in Eq.~(\ref{averagevelocity}) to the experimental velocity $\bar{v}_{exp}(w)$ obtained in ref.~\cite{Ackerman2017}, at $\tau = 0.75$. More technical details are provide in Methods.

In addition to the analytical results on the skyrmion dynamics, we also implemented a numerical approach with a view on future investigation of collective dynamics of interacting skyrmions, and also to account for the effects of temperature which is duscussed below.. To this end, we used an open source {\it Large-scale Atomic/Molecular Massively Parallel Simulator} (LAMMPS) \cite{PLIMPTON19951}. Equations~(\ref{eq_of_motion}) are solved by using Langevin dynamics simulations with Velocity Verlet integration \cite{PhysRev.159.98}. 

Figure~\ref{Fig3}(a) shows typical results for the displacement of a skyrmion as a function of time in response to turning the electric field {\it on} and {\it off}. The lines correspond to Eq.~(\ref{displacement}), and the symbols are the results of the numerical simulations. In agreement with the experimental observations, the skyrmion speed at early times is smaller during the {\it on} state as compared to the early time speed during the {\it off} state. The total displacement (until the skyrmion stops) is however larger on the {\it on} state. Therefore, if the field is turned {\it on} and {\it off} in cycles with a certain duty cycle, the skyrmion will undergo translational motion in a direction depending on the frequency as shown in Fig.~\ref{Fig3}(b). The two top curves correspond to the displacement in $+\hat{x}$ direction, while the bottom one is for the motion in $-\hat{x}$ direction. The inset demonstrates the $\bar{v}(w)$ calculated at $\tau = 0.75$. The velocity is zero for small frequencies and attains the maximum value of $\bar{v}_{max} \approx 0.088$ at $w_{max}\approx 0.003$. The velocity reverses at $w_0\approx 0.173$ and monotonically decreases as the modulating frequency increases until its limiting value $\lim_{w\rightarrow \infty}v(w)=-0.005$. This formal finding is nonphysical and highlights the limitations of the model, which is valid only for low-to-moderate frequencies of the driving electric field. 

Figure~\ref{Fig4}(b) shows average velocity as a function of the frequency, for several values of the duty cycle. The average velocity changes from positive to negative as the frequency increases, provided $\tau$ is not too large, and above a certain threshold, we find the $\bar{v}$ tends to a finite positive value as $w\rightarrow\infty$. This is related to the model limitation, as already mentioned above. We also emphasize that the model implicitly assumes that not only $w$,  but also $1/t_{on}$ and $1/t_{off}$ are not too large. This makes the model inapplicable in the regimes where $\tau$ is close to zero or unity. Another relevant effect of increasing $\tau$ is shifting of both $w_{max}$ and  $\hat{v}_{max}$ to larger values, as shown in Fig.~\ref{Fig4}(b).  

\begin{figure*}[t!]
\includegraphics[width=2.0\columnwidth]{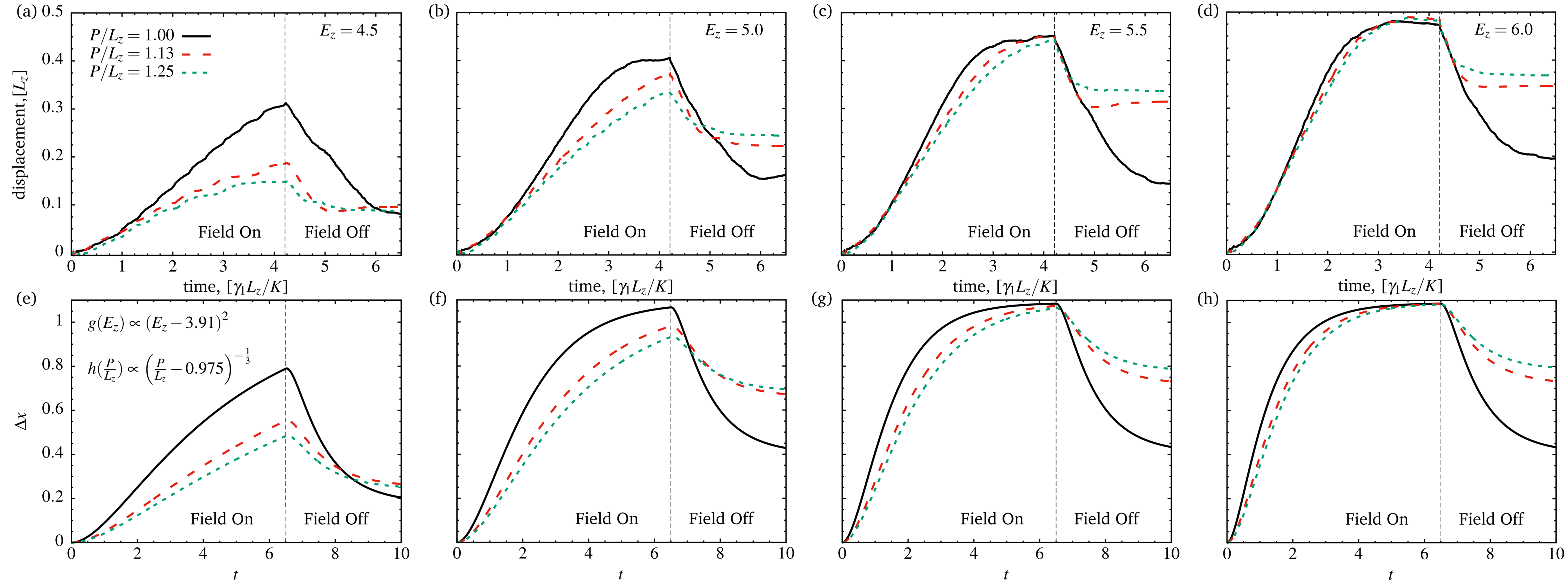} 
  \caption{{\bf Dependence of the model parameters on the electric field and cholesteric pitch.} Skyrmion's displacement as a function of time $t$, where $t=0$ corresponds to switching the electric field {\it on}. At the time marked by the vertical dashed line, the field is switched {\it off} and the skyrmion moves in the opposite direction. Several values of the cholesteric pitch $P$ and electric field strength $E_z$ are considered. (a)-(d) Results of the numerical minimization of the Frank-Oseen free energy. Here $E_z$ is given in units of $\sqrt{K/\varepsilon_0\Delta\varepsilon L_z^2}$, where $K$ is the average elastic constant. (e)-(f) Analytical results obtained by using the function $g(E_z)$ and $h(P/L_z)$ defined in Eqs.~(\ref{params_on}) and (\ref{params_off}). These functions are determined by minimizing the difference between the analytical and numerical ratios of the final displacement when the field is {\it on} to the total displacement.}
  \label{Fig6}
\end{figure*}

Additional influences of $\tau$ on the average velocity are shown in Figs.~\ref{Fig5}(a)-(c), which display colour coded two-dimensional (2D) plots of $\hat{v}$ as a function of $\tau$ (on the vertical axis) and $w$, $|A_{r}|$, and $c_{r}$ on the horizontal axes of panels (a), (b), and (c), respectively. In the figures, the square and $\times$ symbols represent negative and positive velocities, respectively. The solid lines are obtained by setting $\hat{v}=0$ in Eq.~(\ref{averagevelocity}) which gives the following constraint on the model parameters
%
\begin{align}
    A_{r} = 
    c_{r} \frac{e^{- \tau \frac{\alpha_{on}}{w}}-1}
    {1-e^{-\left(1-\tau\right) \frac{\alpha_{on} c_{r}}{w}}},
    \label{eq:zero_condition}
\end{align}
which may be be approximated as follows:
%
\begin{align}
    |A_{r}| \approx 
    \left\{
    \begin{array}{c}
    \quad c_{r}, \quad w \ll 1
    \\
    \frac{\tau}{1-\tau}, \quad w \gg 1
    \end{array}
    \right. .
    \label{Ar}
\end{align}
These limits determine the range of values for parameter $A_{r}$ to ensure that $\bar{v} = 0$, as the frequency is varied. For instance, for the values of $A_{r}$ and $c_{r}$ used in Figs.~\ref{Fig3} and \ref{Fig4}(b), the average velocity will never be zero if $\tau > |A_{r}|/(1+|A_{r}|)$.

The above result also motivates the study of the influence of both $A_{r}$ and $c_{r}$ on the average velocity when the duty is fixed. Figures.~\ref{Fig5}(d)-(f) show the colour coded 2D maps of the average velocity for several pairs of the model parameters, $(A_{r},w)$, in Fig.~\ref{Fig5}(d) and $(c_{r},w)$ in Fig. \ref{Fig5}(e). Fig.~\ref{Fig5}(f) shows the dependence of $A_{r}$ on $c_{r}$ (the solid line) according to Eq.~(\ref{eq:zero_condition}) at $w=0.173$, which is the frequency when $\bar{v} = 0$ in Fig.~\ref{Fig3}.

\vspace{0.5cm}

\noindent {\bf On the dependence of the model parameters on electric field and cholesteric pitch.} The effective displacement of the skyrmion also depends on the strength of the electric field $E_z$ since its dynamics originate from the response to it. Furthermore, changing the cholesteric pitch $P$ implies changing the effective size of the skyrmion, which is also expected to contribute to the skyrmion dynamics. In this section, we discuss the dependence of the model parameters on $E_z$ and $P$ semi-quantitatively. 

Initially, we carried out fine grained analysis by minimizing the Frank-Oseen free energy for several values of $E_z$ and $P$. The corresponding effective skyrmion displacement curves in response to switching the field {\it on} and {\it off} are collected in Figs.~\ref{Fig6}(a)-(d). The dependence of $A'_{on}$ and $\alpha_{on}$ on $E_z$ is expected to be $~E_z^2$. Moreover, in the time regime for which our model is developed both parameters have exactly the same functional form as function of $E_z$. This follows by applying linear stability analysis to the coarse grained dynamics proposed in \cite{Long2021} (see Eqs.~(16) and (17) in \cite{Long2021}). 

The numerical results shown in Figs.~\ref{Fig6}(c)-(d) suggest that the $\Delta x_{on}(t\rightarrow\infty)$ is constant and approximately independent of both $E_z$ and $P$. On the other hand, Eq.~(\ref{displacement}) renders that $\Delta x'_{on}(t\rightarrow\infty) = A'_{on}/\alpha_{on}$ (where $\Delta x'_{on}$ is expressed in units of $m/\gamma^{2}$). Combining these two observations, we conclude that $A'_{on}$ and $\alpha_{on}$ must depend identically on $P$. The results in Figs.~\ref{Fig6}(c)-(d) also demonstrate (approximately) that only $A'_{off}$ and not $\alpha_{off}$ depends on $P$. Recall, that $\alpha_{off}$ and $A'_{off}$ can not depend on $E_z$, because the electric field is zero on this branch of $\Delta x'(t)$.  

Summarizing these observations we propose the following phenomenological expressions 
\begin{align}
    &A'_{on}(E_z,P) = a_{on} g \left( E_{z} \right) h \left( P / L_{z} \right),
    \\
    &\alpha_{on}(E_z,P) = b_{on} g \left( E_{z} \right) h \left( P / L_{z} \right),
    \label{params_on} \\    
    & A'_{off}(P) =A_{r} a_{on} h \left( P / L_{z} \right),
    \label{params_off}
\end{align}
where $a_{on}$ has the same dimension as $A'_{on}$, and $A_{r}$ and $b_{on}$ are dimensionless parameters. For the sake of simplicity, we assume that the dependence on $E_z$ and $P$ can be separated into dimensionless functions $g(E_z)\sim E_z^2$ and $h(P/L_z)$ and we use a single function $h$ to account for the dependence on $P$ of both $A'_{on}$ and $A'_{off}$. Recall, that we also assume that $\alpha_{off}$ does not depend on $P$. 

Figures~\ref{Fig6}(a)-(d) show that by increasing the pitch $|\Delta x_{off}|$ decreases, which means that $A'_{off}(P)$ as a function of $P$ must also decrease. For a sake of simpliciy we assume $h(P/L_z)\sim (P/L_z)^{-z}$ with $z>0$. 

\begin{figure}[t!]
\includegraphics[width=0.9\columnwidth]{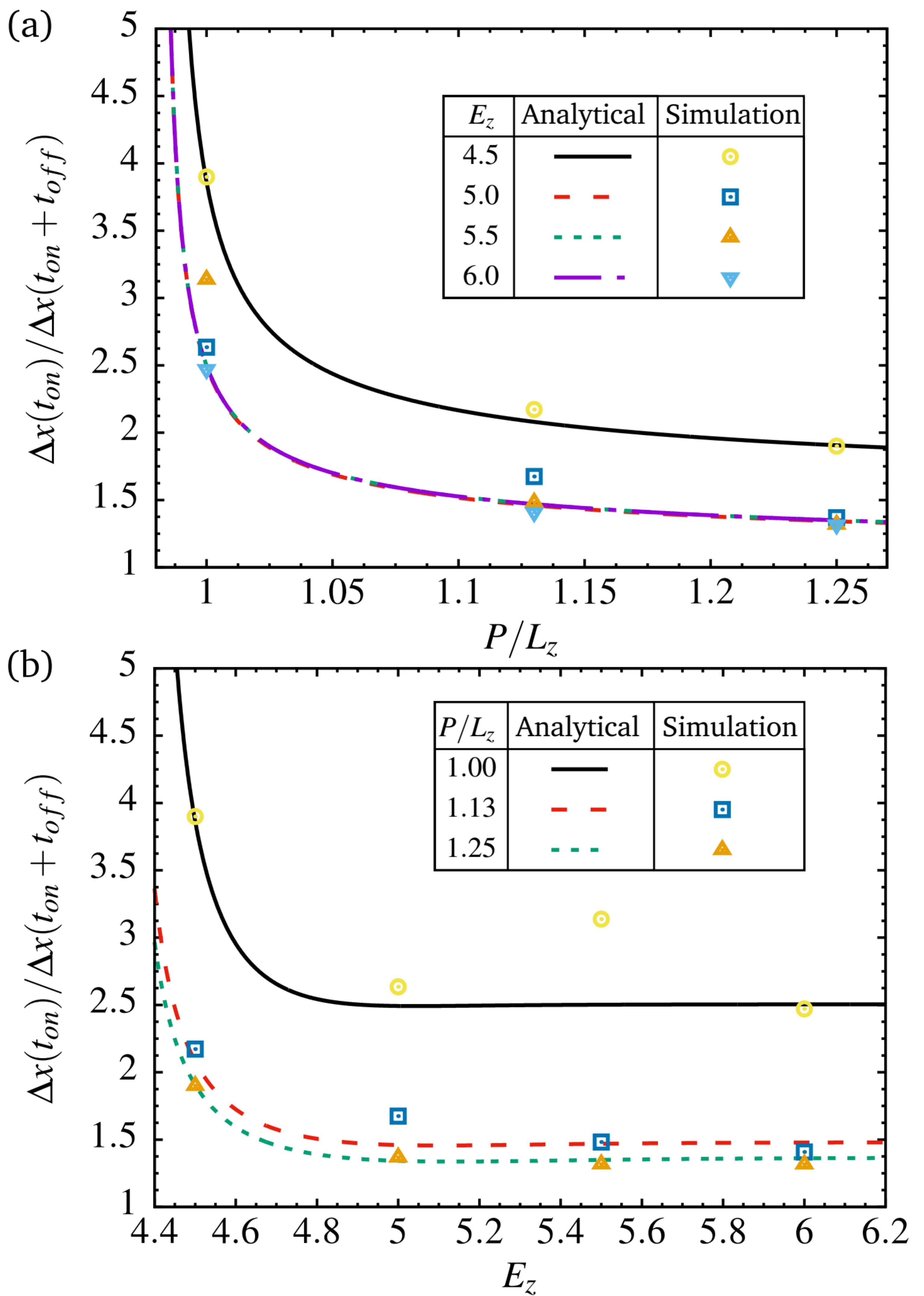} 
  \caption{Ratio of the displacement achieved during the field {\it on} state to the total displacement along the full period as a function of (a) the pitch $P$, and (b) the electric field $E_z$. The later is given in units of $\sqrt{K/\varepsilon_0\Delta\varepsilon L_z^2}$, where $K$ is the average elastic constant. The symbols represent the results of the numerical minimization of the Frank-Oseen free energy, and the lines correspond to analytical results obtained by fitting Eqs.~(\ref{params_on}) and (\ref{params_off}) to the numerical curves in Figs.~\ref{Fig6}(a)-(d).}
  \label{Fig7}
\end{figure}

Returning to dimensionless coordinates units, where now $\Delta x_n$ is expressed in terms of $a_{on} m/\gamma^{2}$, and supported by the above arguments, the analytical displacement is fitted to the computational one. The resulting fitting curves are presented in Figs.~\ref{Fig6}(e)-(h) with 
\begin{align}
    g \left( E_{z} \right) =
    \left( 0.9~E_{z} - 3.52 \right)^{2};
    \\
    h \left( P / L_{z} \right) =
    \left( 50~P / L_{z} - 48.77 \right)^{-1 / 3},
    \hspace*{-1cm}
\end{align}
and where $b_{on}=0.919$; for $\alpha_{off}$ and $A_{r}$ we use the same values as before, i.e., $0.029$ and $-3.02$, respectively. The fitting is performed in such a way that the difference between the analytical and numerical ratios of the displacements $\Delta x (t_{on})/\Delta x(t_{on} + t_{off})$ is minimized. Figure~(\ref{Fig7}) compares the resulting analytical and numerical ratios as functions of the pitch and the field strength.

The above results demonstrate how the coarse grained model, through its parameters, is related to the fine grained one. It is known that the size of the skyrmion is affected by $P/L_z$, becoming smaller and having less free elastic energy when this ratio increases \cite{Sohn2019a}. This could be one of the reasons why $h(P/L_z)$ decreases with $P/L_z$ increasing because as $P/L_{z}$ grows, the effective elastic driving force on the skyrmion gets smaller due to the reduction of the elastic free energy. On the other hand, the tilted orientation of the skyrmion depends on the magnitude of the electric field when it is turned on, tending to orient the far field director in-plane as the magnitude of the electric field increases \cite{Ackerman2017}. In other words, as the field increases, the effective driving force becomes stronger, since the difference between axially symmetric and tilted director textures also increases. This implies that the model's parameters $A'_{on}$ and $\alpha_{on}$ increase when the electric field increases.

\begin{figure}[t!]
\includegraphics[width=0.9\columnwidth]{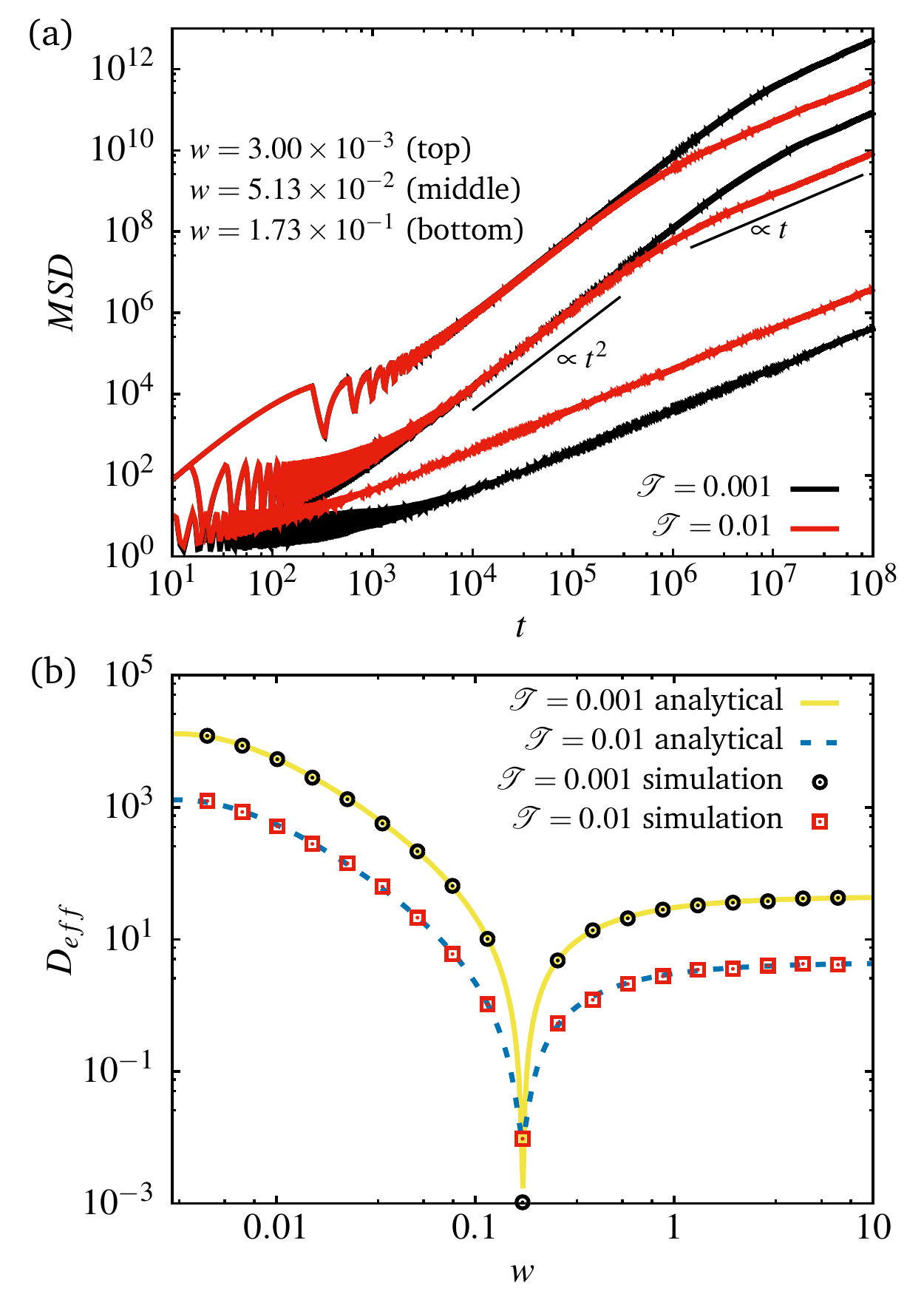} 
  \caption{{\bf Effective diffusion constant.} (a) Mean square displacement for 1000 non-interacting skyrmions as a function of time for several values of  the frequency, and the dimensionless temperature ${\cal{T}} = k_{B} {\cal{T}'} \gamma^2/m A_{on}'^2$, the duty cycle $\tau = 0.75$. (b) Effective  diffusion coefficient $D_{eff}$ as a function of $w$ for several values of ${\cal T}$. $D_{eff}$ is presented in units of $m A_{on}'^2/\gamma^3$, and the symbols are obtained form the long time behaviour $MSD(t) = 4D_{eff}t$. The diffusion constant has a global minimum at $w \approx 0.173$ where the average skyrmion velocity is zero (see Fig.~\ref{Fig4}(b))). Solid and dashed lines correspond to the analytical result of the active Brownian particle model $D_{eff} = {\cal T} + {\bar{v}^{2}}/2D_{R}$, where the dimensionless rotational diffusion constant $D_R = 3 {\cal T}/4 R^2$, at the dimensionless particle radius $R=50$.}
  \label{Fig8}
\end{figure}

\vspace{0.5cm}

\noindent {\bf Effective diffusion coefficient.} Finally, we discuss the effects of finite temperature on the dynamics of skyrmions, also taking into account rotational diffusion. To this end, we calculate, using LAMMPS, the mean square displacement $MSD(t)$ as a function of time for several values of temperature and the frequency of the driving electric field. The resulting curves are shown in Fig.~\ref{Fig8}(a). The oscillatory behaviour at short times $t<10^4$ is due to the change of the field between the {\it on} and {\it off} states, which causes the skyrmion to change its direction of motion as discussed before. We observe a superdiffusive behaviour, when $MSD(t)$ has a term $\propto t^2$, for intermediate times and for the two lowest values of $w$. The superdiffusive regime is also revealed by active Brownian particle model \cite{RevModPhys.88.045006}, which for a 2D system predicts $MSD(t) \approx 4D_T t+2u_a^2t^2$, where $u_a$ is the particle self-propulsion velocity and $D_T$ is the diffusion coefficient for the passive case with $u_a = 0$. Finally for $t\sim 1/D_R$ the $MSD(t)$ crosses over to a diffusive regime where $MSD(t) = D_{eff} t$, $D_R$ above is the rotational diffusion constant. The two bottom curves in Fig.~\ref{Fig8}(a) correspond to the frequency at which the average skyrmion velocity is zero. Therefore, in this case, the superdiffusive regime is not observed.

Figure~\ref{Fig8}(b) shows the long time diffusion coefficient $D_{eff}$ of the skyrmion as a function of the frequency $w$. The frequency at which $D_{eff}$ attains minimum corresponds to zero average velocity of the skyrmion. It is known that for active Brownian particles $D_{eff} = D_{T} + \frac{u_a^2}{2D_{R}}$ \cite{RevModPhys.88.045006}. We evaluate $D_T$ and $D_R$ for the active skyrmions by using Stokes-Einstein equations and with the friction coefficient of a spherical colloidal particle of radius $R$. We also replace the self-propulsion velocity in the expression for $D_{eff}$ above with the skyrmion's average velocity, $\bar{v}$ from Fig.~\ref{Fig4}(b). The resulting theoretical predictions for $D_{eff}(w)$ are plotted as lines in Fig.~\ref{Fig8}(b). The theoretical predictions perfectly match numerical data calculated from the late time $MSD(t)$. This observation shows that the active skyrmions behave as effective active Brownian particles with the self-propulsion velocity equal to the skyrmion average velocity  $\bar{v}$.
\vspace{0.5cm}

\noindent {\large \bf Discussion}

\noindent We have studied the translation of LC skyrmions in response to the time dependent electric field. Starting from the fine grained description of the underlying LC structure using the Frank-Oseen elastic free energy, we develop a coarse grained model where the coordinates of the skyrmion center of mass are the only degrees of freedom. The non-reciprocal morphing of the LC director field in response to a changing electric field is accounted for by time dependent effective forces acting upon the skyrmion center. The functional form of the forces is dictated by the behaviour of the skyrmion velocity (overdamped regime) as per fine grained Frank-Oseen description. In the vicinity of the steady states, the velocity decays with time exponentially, where the relaxation times are controlled by the amplitude of the electric field and the anchoring strength at the confining walls. 

We have also analysed the dependence of the phenomenological parameters of the coarse grained model on the cholesteric pitch and the field strength. We have obtained approximate analytical expressions for the parameters by fitting coarse grained skyrmion trajectories to the trajectories calculated via numerical minimization of the Frank-Oseen model. 

The proposed model is limited to low frequencies of the driving electric fields, and also to long enough duration of the field-{\it on} and field-{\it off} states, such that the skyrmion completely relaxes towards its equilibrium configurations. This limitation may be partially overcome, without changing the functional forms of the effective forces, by assigning to each value of the force a corresponding instantaneous skyrmion configuration. Next, upon switching the field between the {\it on} and {\it off} states, the new force must be initiated at that time, which corresponds to the latest skyrmion configuration. At present, the forces are always initiated at time zero, meaning that the skyrmion has relaxed completely. 

Our model captures semi-quantitatively the main features of the skyrmion average velocity as a function of the frequency and the duty cycle. The velocity attains the global maximum at relatively low values of the frequency, which both shift towards larger values as the duty cycle increases. We have also proposed simple empirical expressions for the dependence of the phenomenological model parameters on the material parameters (the cholesteric pitch) and the amplitude of the electric field.

The proposed coarse grained model is suitable for particle-based simulations of non-interacting skyrmions, and could be straightforwardly extended by adding effective skyrmion interactions determined from the fine grained Frank-Oseen model. Both static and out-of-equilibrium forces are possible in this case, as evidenced by experimental studies \cite{Sohn2018,Sohn2020,Sohn2019a,Sohn2019}.   
\vspace{0.5cm}

\noindent {\large \bf Methods}

\noindent{\bf Fixing free model parameters by comparison with the experimental result}. Analysis of Eq.~(\ref{averagevelocity}) shows that $\bar{v}(w)$ has the global maximum at $w_{max}$, crosses zero at  $w=w_0$, and has the global minimum  $\bar{v}_{min} = \lim_{w\rightarrow \infty}\bar{v}(w)<0$. The last result highlights the limitation of the model, which by design is applicable for small-to-moderate frequencies only. On physical grounds it is expected that the average velocity goes to zero as frequency increases, because of the finite relaxation and response times of the nematic director.

We use the experimental curve $\bar{v}_{exp}(w)$, which also exhibits the global maximum, zero point, and a global minimum at finite $w$, to formulate the following two conditions: i) the ratio $\bar{v}_{max}/\bar{v}_{min}$ of the maximal to minimal average velocity in the model equals analogous ratio $\bar{v}_{exp,max}/\bar{v}_{exp,min}$ in the experiments; ii) $w_{0}/w_{max}$ is equal to the experimental ratio $w_{exp,0}/w_{exp,max}$. With the conditions i) and ii) the model has only one free parameter, for which we select $c_r$ because it may easily be estimated from the experimental skyrmion displacement curve in \cite{Ackerman2017} giving $c_r\approx 4.50$. With this we find $A_{r} = - 3.02$ and $\alpha_{on}=6.4 \times 10^{-3}$. These values are used throughout the rest of the paper, with few exceptions which are explicitly stated. 
\vspace{0.5cm}

\noindent{\bf Numerical minimization of the Frank-Oseen elastic free energy.} We use the Frank-Oseen elastic free energy to study the 3D director configurations in response to pulse width moduleted electric fields. The free energy for chiral nematic LCs can be expressed as follows:
\begin{equation}
    f_{el}=\frac{1}{2}K_1(\nabla\cdot\nvec)^2+\frac{1}{2}K_2(\nvec\cdot\nabla\times\nvec-\frac{2\pi}{P})^2+\frac{1}{2}K_3(\nvec\times\nabla\times\nvec)^2
    \label{eq:frank_oseen}
\end{equation}
where $\nvec$ is the nematic director, $K_1, K_{2}$ and $K_3$ are positive elastic constants describing splay, twist and bend director distortions, respectively; $P$ is the cholesteric pitch with a defined length at which the director twist by $2\pi$. External electric field $\Evec$ couples to the nematic director according to the following Electric free energy density 
\begin{equation} 
    f_{E}= \frac{1}{2} \varepsilon_0\Delta\varepsilon (\Evec \cdot \nvec)^2.
    \label{eq:electric_energy}
\end{equation}
Here, $\varepsilon_0$ is the Vacuum permittivity and $\Delta\varepsilon$ is the dielectric anisotropy. We emphasize that Eq.~(\ref{eq:electric_energy}) approximates the local electric field by the constant external field. This approximation is valid only for weak fields, as used in the experiments \cite{Ackerman2017,Sohn2020}. 
The relaxation dynamics of the director field is governed by the following equation:
\begin{equation}
 \partial_t n_\mu = -\frac{1}{\gamma_1} \frac{\delta }{\delta n_\mu} \int \Big(f_{el} + f_{E}\Big)dV ,
 \label{director-time-eq}
\end{equation}
where $\gamma_1$ is the rotational viscosity, and integration is carried out over the three dimensional (3D) domain $V$ occupied by LC. We assume $V=L\times L\times L_z$, where $L_z$ is the separation between the confining rigid surfaces of the area $L \times L$. The director obeys rigid homeotropic boundary conditions at the surfaces, and we apply periodic boundary conditions in the $x$ and $y$ directions perpendicular to the surfaces. Equation (\ref{director-time-eq}) must be solve subject to the constraint $(\nvec \cdot \nvec) = 1$. The spatial derivatives on the r.h.s of Eq.~(\ref{director-time-eq}) are approximated by using finite-differences and the integration over time is performed using the fourth-order Runge-Kutta method. The values of the model parameters used in this study are provided in table \ref{table}.

\begin{table}[t!]
\caption{\label{tab1} Parameters used in numerical integration of Eq.~(\ref{director-time-eq}) and physical units.}
\footnotesize
\begin{tabular}{@{}|l|l|l|l|}
\hline
symbol&sim. units & physical units&description\\
\hline
$\Delta x$&1 & 0.3125 $\mu$m& lattice spacing\\
$\Delta t$&1 & 10$^{-6}$ s& time step\\
$L$&300 & $\approx 94 \mu m$ & simulation box side length\\
$L_z$&33 & $10 \mu m$ & confining surfaces separation\\
$K_{1}$&17.2 &17.2$\times 10^{-12}$ N & splay elastic constant\\
$K_{2}$&7.51 &7.51$\times 10^{-12}$ N & twist elastic constant\\
$K_{3}$&17.2 &17.2$\times 10^{-12}$ N & bend elastic constant\\
$\gamma_1$&162&0.162 Pa s&director rotational viscosity\\
$\Delta\varepsilon$ & -3.7 & -3.7& dielectric anisotropy\\
\hline
\end{tabular}\\
\label{table}
\end{table}
\normalsize

As an initial conditions for numerical integration of Eq.~(\ref{director-time-eq}), we use simple axially symmetric Ansatz for the director field: 
\begin{align}
 &n_x = \sin(a) \sin(mb + g)\nonumber\\
 &n_y = \sin(a) \cos(mb + g)\nonumber\\
 &n_z = -\cos(a),
 \label{toron_ansatz}
\end{align}
where (for $z' \equiv z-C_z $)
\begin{equation}
 a =  
 \begin{cases}
\frac{\pi}{2}\left[1 - \tanh\left(\frac{B}{2}(r-\sqrt{R^2-(z')^2})\right)\right],R>|z'|\\
0, R\leq|z'|
\end{cases}
\end{equation}
\begin{eqnarray}
 && b = \tan^{-1}\left(\frac{x-C_x}{y-C_y}\right)\\
 && r=\sqrt{(x-C_x)^2+(y-C_y)^2} .
\end{eqnarray}
Here $(C_x,C_y,C_z)^{\mathrm{T}}$ defines the coordinates of the Ansatz center, $R$ controls the size of the skyrmion, $B$ controls the width of the twisted wall of the skyrmionic tube, $m$ is the winding number of the skyrmion, $g$ controls the direction of the skyrmion, $r$ is the distance from a given point $(x,y,z)^{\mathrm{T}}$ to the skyrmion symmetry axis, and $b$ is the $2D$ polar angle. The values of the parameters used in the simulation are (in simulation units): $m=1$, $g = \pi/2$, $R=0.45L_z$, $B=0.5$, $C_x=C_y=L/2$, $C_z=H/2$.

The director field with embedded skyrmion configuration (Fig.~\ref{fig_configs_FO_model}(a)) will morph in response to time dependent electric field, and the spatial localisation of the skyrmion will change, resembling an effective $2D$ translational motion, recall that the skyrmion is squeezed between two plates. To characterise this in-plane motion we define the skyrmion's ``center of mass'' $(x_c,y_c)^{\mathrm{T}}$ as:
\begin{equation}
(x_c,y_c)^{\mathrm{T}} = \frac{\int^*n_z(\mathbf{r})(x,y)^{\mathrm{T}}dV}{\int^*n_z(\mathbf{r})dV},  
\label{toron-center-of-mass}
\end{equation}
 where the integration is over domain defined by $n_z({\mathbf{r}})>0.3$, which corresponds to the preimage (the red region in Fig.~\ref{fig_configs_FO_model}(a)) of a spherical cap at the north pole on the order parameter space $\mathrm{S}^2$ of the vectorised $\nvec$.

 Before applying an external electric field $(0,0,E_z)^{\mathrm{T}}$, we must brake ``by hand'' the axial symmetry of the {\it Ansazt} in Eq.~(\ref{toron_ansatz}). 
 To this end, we first set $\Evec=(0,E_0,E_0)^{\mathrm{T}}$, and carry out 500 Runge-Kutta iterations on discretised Eq.~(\ref{director-time-eq}). This tilts the far field director along $x-$axis. Next, we switch the field off and relax the director field for 5500 time steps, the resulting skyrmion configuration will remain slightly asymmetric, which is crucial to achieve a net displacement of the skyrmion in subsequent field cycling. After these initial 6000 time step the skyrmion center attains some new position $x_c=X_0\neq C_x$ on the $x-$axis. The $y$ coordinate of the skyrmion center $y_c$ stays close to $C_y$. 
 
 After this initial stage, we proceed to the production runs. We set $\Evec=(0,0,E_z)^{\mathrm{T}}$ and relax $\nvec$ according to Eq.~(\ref{director-time-eq}) until the equilibrium state is reached. 
\vspace{0.5cm}

\noindent {\large \bf Acknowledgments}

\noindent We acknowledge financial support from the Portuguese Foundation for Science and Technology (FCT) under Contracts no. PTDC/FIS-MAC/5689/2020, EXPL/FIS-MAC/0406/2021, CEECIND/00586/2017, UIDB/00618/2020,  and UIDP/00618/2020.

\vspace{0.5cm}

\noindent {\large \bf References}

\bibliography{rsc} 

\begin{thebibliography}{32}%
\makeatletter
\providecommand \@ifxundefined [1]{%
 \@ifx{#1\undefined}
}%
\providecommand \@ifnum [1]{%
 \ifnum #1\expandafter \@firstoftwo
 \else \expandafter \@secondoftwo
 \fi
}%
\providecommand \@ifx [1]{%
 \ifx #1\expandafter \@firstoftwo
 \else \expandafter \@secondoftwo
 \fi
}%
\providecommand \natexlab [1]{#1}%
\providecommand \enquote  [1]{``#1''}%
\providecommand \bibnamefont  [1]{#1}%
\providecommand \bibfnamefont [1]{#1}%
\providecommand \citenamefont [1]{#1}%
\providecommand \href@noop [0]{\@secondoftwo}%
\providecommand \href [0]{\begingroup \@sanitize@url \@href}%
\providecommand \@href[1]{\@@startlink{#1}\@@href}%
\providecommand \@@href[1]{\endgroup#1\@@endlink}%
\providecommand \@sanitize@url [0]{\catcode `\\12\catcode `\$12\catcode
  `\&12\catcode `\#12\catcode `\^12\catcode `\_12\catcode `\%12\relax}%
\providecommand \@@startlink[1]{}%
\providecommand \@@endlink[0]{}%
\providecommand \url  [0]{\begingroup\@sanitize@url \@url }%
\providecommand \@url [1]{\endgroup\@href {#1}{\urlprefix }}%
\providecommand \urlprefix  [0]{URL }%
\providecommand \Eprint [0]{\href }%
\providecommand \doibase [0]{https://doi.org/}%
\providecommand \selectlanguage [0]{\@gobble}%
\providecommand \bibinfo  [0]{\@secondoftwo}%
\providecommand \bibfield  [0]{\@secondoftwo}%
\providecommand \translation [1]{[#1]}%
\providecommand \BibitemOpen [0]{}%
\providecommand \bibitemStop [0]{}%
\providecommand \bibitemNoStop [0]{.\EOS\space}%
\providecommand \EOS [0]{\spacefactor3000\relax}%
\providecommand \BibitemShut  [1]{\csname bibitem#1\endcsname}%
\let\auto@bib@innerbib\@empty
\bibitem [{\citenamefont {Smalyukh}\ \emph {et~al.}(2010)\citenamefont
  {Smalyukh}, \citenamefont {Lansac}, \citenamefont {Clark},\ and\
  \citenamefont {Trivedi}}]{Smalyukh2010}%
  \BibitemOpen
  \bibfield  {author} {\bibinfo {author} {\bibfnamefont {I.~I.}\ \bibnamefont
  {Smalyukh}}, \bibinfo {author} {\bibfnamefont {Y.}~\bibnamefont {Lansac}},
  \bibinfo {author} {\bibfnamefont {N.~A.}\ \bibnamefont {Clark}},\ and\
  \bibinfo {author} {\bibfnamefont {R.~P.}\ \bibnamefont {Trivedi}},\
  }\bibfield  {title} {\bibinfo {title} {Three-dimensional structure and
  multistable optical switching of triple-twisted particle-like excitations in
  anisotropic fluids},\ }\href {https://www.nature.com/articles/nmat2592}
  {\bibfield  {journal} {\bibinfo  {journal} {Nat. Mater}\ }\textbf {\bibinfo
  {volume} {9}} (\bibinfo {year} {2010})}\BibitemShut {NoStop}%
\bibitem [{\citenamefont {Ackerman}\ \emph {et~al.}(2014)\citenamefont
  {Ackerman}, \citenamefont {Trivedi}, \citenamefont {Senyuk}, \citenamefont
  {van~de Lagemaat},\ and\ \citenamefont {Smalyukh}}]{Ackerman2014}%
  \BibitemOpen
  \bibfield  {author} {\bibinfo {author} {\bibfnamefont {P.~J.}\ \bibnamefont
  {Ackerman}}, \bibinfo {author} {\bibfnamefont {R.~P.}\ \bibnamefont
  {Trivedi}}, \bibinfo {author} {\bibfnamefont {B.}~\bibnamefont {Senyuk}},
  \bibinfo {author} {\bibfnamefont {J.}~\bibnamefont {van~de Lagemaat}},\ and\
  \bibinfo {author} {\bibfnamefont {I.~I.}\ \bibnamefont {Smalyukh}},\
  }\bibfield  {title} {\bibinfo {title} {Two-dimensional skyrmions and other
  solitonic structures in confinement-frustrated chiral nematics},\ }\href
  {https://link.aps.org/doi/10.1103/PhysRevE.90.012505} {\bibfield  {journal}
  {\bibinfo  {journal} {Phys. Rev. E}\ }\textbf {\bibinfo {volume} {90}},\
  \bibinfo {pages} {012505} (\bibinfo {year} {2014})}\BibitemShut {NoStop}%
\bibitem [{\citenamefont {Ackerman}\ \emph {et~al.}(2017)\citenamefont
  {Ackerman}, \citenamefont {Boyle},\ and\ \citenamefont
  {Smalyukh}}]{Ackerman2017}%
  \BibitemOpen
  \bibfield  {author} {\bibinfo {author} {\bibfnamefont {P.~J.}\ \bibnamefont
  {Ackerman}}, \bibinfo {author} {\bibfnamefont {T.}~\bibnamefont {Boyle}},\
  and\ \bibinfo {author} {\bibfnamefont {I.~I.}\ \bibnamefont {Smalyukh}},\
  }\bibfield  {title} {\bibinfo {title} {Squirming motion of baby skyrmions in
  nematic fluids},\ }\href {https://www.nature.com/articles/s41467-017-00659-5}
  {\bibfield  {journal} {\bibinfo  {journal} {Nat. Commun.}\ }\textbf {\bibinfo
  {volume} {8}} (\bibinfo {year} {2017})}\BibitemShut {NoStop}%
\bibitem [{\citenamefont {Sohn}\ \emph {et~al.}(2018)\citenamefont {Sohn},
  \citenamefont {Ackerman}, \citenamefont {Boyle}, \citenamefont {Sheetah},
  \citenamefont {Fornberg},\ and\ \citenamefont {Smalyukh}}]{Sohn2018}%
  \BibitemOpen
  \bibfield  {author} {\bibinfo {author} {\bibfnamefont {H.~R.}\ \bibnamefont
  {Sohn}}, \bibinfo {author} {\bibfnamefont {P.~J.}\ \bibnamefont {Ackerman}},
  \bibinfo {author} {\bibfnamefont {T.~J.}\ \bibnamefont {Boyle}}, \bibinfo
  {author} {\bibfnamefont {G.~H.}\ \bibnamefont {Sheetah}}, \bibinfo {author}
  {\bibfnamefont {B.}~\bibnamefont {Fornberg}},\ and\ \bibinfo {author}
  {\bibfnamefont {I.~I.}\ \bibnamefont {Smalyukh}},\ }\bibfield  {title}
  {\bibinfo {title} {Dynamics of topological solitons, knotted streamlines, and
  transport of cargo in liquid crystals},\ }\href
  {https://journals.aps.org/pre/abstract/10.1103/PhysRevE.97.052701} {\bibfield
   {journal} {\bibinfo  {journal} {Phys. Rev. E}\ }\textbf {\bibinfo {volume}
  {97}} (\bibinfo {year} {2018})}\BibitemShut {NoStop}%
\bibitem [{\citenamefont {Sohn}\ \emph
  {et~al.}(2019{\natexlab{a}})\citenamefont {Sohn}, \citenamefont {Liu},
  \citenamefont {Wang},\ and\ \citenamefont {Smalyukh}}]{Sohn2019a}%
  \BibitemOpen
  \bibfield  {author} {\bibinfo {author} {\bibfnamefont {H.~R.~O.}\
  \bibnamefont {Sohn}}, \bibinfo {author} {\bibfnamefont {C.~D.}\ \bibnamefont
  {Liu}}, \bibinfo {author} {\bibfnamefont {Y.}~\bibnamefont {Wang}},\ and\
  \bibinfo {author} {\bibfnamefont {I.~I.}\ \bibnamefont {Smalyukh}},\
  }\bibfield  {title} {\bibinfo {title} {Light-controlled skyrmions and torons
  as reconfigurable particles},\ }\href
  {https://opg.optica.org/oe/abstract.cfm?URI=oe-27-20-29055} {\bibfield
  {journal} {\bibinfo  {journal} {Opt. Express}\ }\textbf {\bibinfo {volume}
  {27}},\ \bibinfo {pages} {29055} (\bibinfo {year}
  {2019}{\natexlab{a}})}\BibitemShut {NoStop}%
\bibitem [{\citenamefont {Sohn}\ \emph
  {et~al.}(2019{\natexlab{b}})\citenamefont {Sohn}, \citenamefont {Liu},\ and\
  \citenamefont {Smalyukh}}]{Sohn2019}%
  \BibitemOpen
  \bibfield  {author} {\bibinfo {author} {\bibfnamefont {H.~R.}\ \bibnamefont
  {Sohn}}, \bibinfo {author} {\bibfnamefont {C.~D.}\ \bibnamefont {Liu}},\ and\
  \bibinfo {author} {\bibfnamefont {I.~I.}\ \bibnamefont {Smalyukh}},\
  }\bibfield  {title} {\bibinfo {title} {Schools of skyrmions with electrically
  tunable elastic interactions},\ }\href
  {https://www.nature.com/articles/s41467-019-12723-3} {\bibfield  {journal}
  {\bibinfo  {journal} {Nat. Commun.}\ }\textbf {\bibinfo {volume} {10}}
  (\bibinfo {year} {2019}{\natexlab{b}})}\BibitemShut {NoStop}%
\bibitem [{\citenamefont {Song}\ \emph {et~al.}(2021)\citenamefont {Song},
  \citenamefont {Kerber}, \citenamefont {Rothörl}, \citenamefont {Ge},
  \citenamefont {Raab}, \citenamefont {Seng}, \citenamefont {Brems},
  \citenamefont {Dittrich}, \citenamefont {Reeve}, \citenamefont {Wang},
  \citenamefont {Liu}, \citenamefont {Virnau},\ and\ \citenamefont
  {Kläui}}]{Song2021}%
  \BibitemOpen
  \bibfield  {author} {\bibinfo {author} {\bibfnamefont {C.}~\bibnamefont
  {Song}}, \bibinfo {author} {\bibfnamefont {N.}~\bibnamefont {Kerber}},
  \bibinfo {author} {\bibfnamefont {J.}~\bibnamefont {Rothörl}}, \bibinfo
  {author} {\bibfnamefont {Y.}~\bibnamefont {Ge}}, \bibinfo {author}
  {\bibfnamefont {K.}~\bibnamefont {Raab}}, \bibinfo {author} {\bibfnamefont
  {B.}~\bibnamefont {Seng}}, \bibinfo {author} {\bibfnamefont {M.~A.}\
  \bibnamefont {Brems}}, \bibinfo {author} {\bibfnamefont {F.}~\bibnamefont
  {Dittrich}}, \bibinfo {author} {\bibfnamefont {R.~M.}\ \bibnamefont {Reeve}},
  \bibinfo {author} {\bibfnamefont {J.}~\bibnamefont {Wang}}, \bibinfo {author}
  {\bibfnamefont {Q.}~\bibnamefont {Liu}}, \bibinfo {author} {\bibfnamefont
  {P.}~\bibnamefont {Virnau}},\ and\ \bibinfo {author} {\bibfnamefont
  {M.}~\bibnamefont {Kläui}},\ }\bibfield  {title} {\bibinfo {title}
  {Commensurability between element symmetry and the number of skyrmions
  governing skyrmion diffusion in confined geometries},\ }\href
  {https://onlinelibrary.wiley.com/doi/full/10.1002/adfm.202010739} {\bibfield
  {journal} {\bibinfo  {journal} {Adv. Funct. Mater.}\ }\textbf {\bibinfo
  {volume} {31}} (\bibinfo {year} {2021})}\BibitemShut {NoStop}%
\bibitem [{\citenamefont {Bogdanov}\ \emph {et~al.}(2003)\citenamefont
  {Bogdanov}, \citenamefont {R{\"o}\ss{ler}},\ and\ \citenamefont
  {Shestakov}}]{Bogdanov2003}%
  \BibitemOpen
  \bibfield  {author} {\bibinfo {author} {\bibfnamefont {A.~N.}\ \bibnamefont
  {Bogdanov}}, \bibinfo {author} {\bibfnamefont {U.~K.}\ \bibnamefont
  {R{\"o}\ss{ler}}},\ and\ \bibinfo {author} {\bibfnamefont {A.~A.}\
  \bibnamefont {Shestakov}},\ }\bibfield  {title} {\bibinfo {title} {Skyrmions
  in nematic liquid crystals},\ }\href
  {https://link.aps.org/doi/10.1103/PhysRevE.67.016602} {\bibfield  {journal}
  {\bibinfo  {journal} {Phys. Rev. E}\ }\textbf {\bibinfo {volume} {67}},\
  \bibinfo {pages} {016602} (\bibinfo {year} {2003})}\BibitemShut {NoStop}%
\bibitem [{\citenamefont {Duzgun}\ \emph
  {et~al.}(2018{\natexlab{a}})\citenamefont {Duzgun}, \citenamefont
  {Selinger},\ and\ \citenamefont {Saxena}}]{Duzgun2018}%
  \BibitemOpen
  \bibfield  {author} {\bibinfo {author} {\bibfnamefont {A.}~\bibnamefont
  {Duzgun}}, \bibinfo {author} {\bibfnamefont {J.~V.}\ \bibnamefont
  {Selinger}},\ and\ \bibinfo {author} {\bibfnamefont {A.}~\bibnamefont
  {Saxena}},\ }\bibfield  {title} {\bibinfo {title} {Comparing skyrmions and
  merons in chiral liquid crystals and magnets},\ }\href
  {https://doi.org/10.1103/PhysRevE.97.062706} {\bibfield  {journal} {\bibinfo
  {journal} {Phys. Rev. E}\ }\textbf {\bibinfo {volume} {97}},\ \bibinfo
  {pages} {062706} (\bibinfo {year} {2018}{\natexlab{a}})}\BibitemShut
  {NoStop}%
\bibitem [{\citenamefont {Duzgun}\ \emph {et~al.}(2021)\citenamefont {Duzgun},
  \citenamefont {Saxena},\ and\ \citenamefont {Selinger}}]{Duzgun2021}%
  \BibitemOpen
  \bibfield  {author} {\bibinfo {author} {\bibfnamefont {A.}~\bibnamefont
  {Duzgun}}, \bibinfo {author} {\bibfnamefont {A.}~\bibnamefont {Saxena}},\
  and\ \bibinfo {author} {\bibfnamefont {J.~V.}\ \bibnamefont {Selinger}},\
  }\bibfield  {title} {\bibinfo {title} {Alignment-induced reconfigurable walls
  for patterning and assembly of liquid crystal skyrmions},\ }\href@noop {}
  {\bibfield  {journal} {\bibinfo  {journal} {Phys. Rev. Res.}\ }\textbf
  {\bibinfo {volume} {3}} (\bibinfo {year} {2021})}\BibitemShut {NoStop}%
\bibitem [{\citenamefont {Duzgun}\ \emph {et~al.}(2022)\citenamefont {Duzgun},
  \citenamefont {Nisoli}, \citenamefont {Reichhardt},\ and\ \citenamefont
  {Reichhardt}}]{Duzgun2022}%
  \BibitemOpen
  \bibfield  {author} {\bibinfo {author} {\bibfnamefont {A.}~\bibnamefont
  {Duzgun}}, \bibinfo {author} {\bibfnamefont {C.}~\bibnamefont {Nisoli}},
  \bibinfo {author} {\bibfnamefont {C.~J.~O.}\ \bibnamefont {Reichhardt}},\
  and\ \bibinfo {author} {\bibfnamefont {C.}~\bibnamefont {Reichhardt}},\
  }\bibfield  {title} {\bibinfo {title} {Directed motion of liquid crystal
  skyrmions with oscillating fields},\ }\href
  {https://iopscience.iop.org/article/10.1088/1367-2630/ac58b8} {\bibfield
  {journal} {\bibinfo  {journal} {New J. Phys.}\ }\textbf {\bibinfo {volume}
  {24}},\ \bibinfo {pages} {033033} (\bibinfo {year} {2022})}\BibitemShut
  {NoStop}%
\bibitem [{\citenamefont {Coelho}\ \emph {et~al.}(2022)\citenamefont {Coelho},
  \citenamefont {Tasinkevych},\ and\ \citenamefont {Gama}}]{Coelho2022}%
  \BibitemOpen
  \bibfield  {author} {\bibinfo {author} {\bibfnamefont {R.~C.}\ \bibnamefont
  {Coelho}}, \bibinfo {author} {\bibfnamefont {M.}~\bibnamefont
  {Tasinkevych}},\ and\ \bibinfo {author} {\bibfnamefont {M.~M. T.~D.}\
  \bibnamefont {Gama}},\ }\bibfield  {title} {\bibinfo {title} {Dynamics of
  flowing 2d skyrmions},\ }\href@noop {} {\bibfield  {journal} {\bibinfo
  {journal} {J. Phys. Cond. Mat.}\ }\textbf {\bibinfo {volume} {34}} (\bibinfo
  {year} {2022})}\BibitemShut {NoStop}%
\bibitem [{\citenamefont {Long}\ and\ \citenamefont
  {Selinger}(2021)}]{Long2021}%
  \BibitemOpen
  \bibfield  {author} {\bibinfo {author} {\bibfnamefont {C.}~\bibnamefont
  {Long}}\ and\ \bibinfo {author} {\bibfnamefont {J.~V.}\ \bibnamefont
  {Selinger}},\ }\bibfield  {title} {\bibinfo {title} {Coarse-grained theory
  for motion of solitons and skyrmions in liquid crystals},\ }\href@noop {}
  {\bibfield  {journal} {\bibinfo  {journal} {Soft Matter}\ }\textbf {\bibinfo
  {volume} {17}} (\bibinfo {year} {2021})}\BibitemShut {NoStop}%
\bibitem [{\citenamefont {Skyrme}(1962)}]{Skyrme1962}%
  \BibitemOpen
  \bibfield  {author} {\bibinfo {author} {\bibfnamefont {T.}~\bibnamefont
  {Skyrme}},\ }\bibfield  {title} {\bibinfo {title} {A unified field theory of
  mesons and baryons},\ }\href {https://doi.org/10.1016/0029-5582(62)90775-7}
  {\bibfield  {journal} {\bibinfo  {journal} {Nucl. Phys.}\ }\textbf {\bibinfo
  {volume} {31}},\ \bibinfo {pages} {556} (\bibinfo {year} {1962})}\BibitemShut
  {NoStop}%
\bibitem [{\citenamefont {Mühlbauer}\ \emph {et~al.}(2009)\citenamefont
  {Mühlbauer}, \citenamefont {Binz}, \citenamefont {Jonietz}, \citenamefont
  {Pfleiderer}, \citenamefont {Rosch}, \citenamefont {Neubauer}, \citenamefont
  {Georgii},\ and\ \citenamefont {Böni}}]{Muhlbauer:2009}%
  \BibitemOpen
  \bibfield  {author} {\bibinfo {author} {\bibfnamefont {S.}~\bibnamefont
  {Mühlbauer}}, \bibinfo {author} {\bibfnamefont {B.}~\bibnamefont {Binz}},
  \bibinfo {author} {\bibfnamefont {F.}~\bibnamefont {Jonietz}}, \bibinfo
  {author} {\bibfnamefont {C.}~\bibnamefont {Pfleiderer}}, \bibinfo {author}
  {\bibfnamefont {A.}~\bibnamefont {Rosch}}, \bibinfo {author} {\bibfnamefont
  {A.}~\bibnamefont {Neubauer}}, \bibinfo {author} {\bibfnamefont
  {R.}~\bibnamefont {Georgii}},\ and\ \bibinfo {author} {\bibfnamefont
  {P.}~\bibnamefont {Böni}},\ }\bibfield  {title} {\bibinfo {title} {Skyrmion
  lattice in a chiral magnet},\ }\href
  {https://doi.org/10.1126/science.1166767} {\bibfield  {journal} {\bibinfo
  {journal} {Science}\ }\textbf {\bibinfo {volume} {323}},\ \bibinfo {pages}
  {915} (\bibinfo {year} {2009})},\ \Eprint
  {https://arxiv.org/abs/https://www.science.org/doi/pdf/10.1126/science.1166767}
  {https://www.science.org/doi/pdf/10.1126/science.1166767} \BibitemShut
  {NoStop}%
\bibitem [{\citenamefont {Yu}\ \emph {et~al.}(2010)\citenamefont {Yu},
  \citenamefont {Onose}, \citenamefont {Kanazawa}, \citenamefont {Park},
  \citenamefont {Han}, \citenamefont {Matsui}, \citenamefont {Nagaosa},\ and\
  \citenamefont {Tokura}}]{Yu2010}%
  \BibitemOpen
  \bibfield  {author} {\bibinfo {author} {\bibfnamefont {X.~Z.}\ \bibnamefont
  {Yu}}, \bibinfo {author} {\bibfnamefont {Y.}~\bibnamefont {Onose}}, \bibinfo
  {author} {\bibfnamefont {N.}~\bibnamefont {Kanazawa}}, \bibinfo {author}
  {\bibfnamefont {J.~H.}\ \bibnamefont {Park}}, \bibinfo {author}
  {\bibfnamefont {J.~H.}\ \bibnamefont {Han}}, \bibinfo {author} {\bibfnamefont
  {Y.}~\bibnamefont {Matsui}}, \bibinfo {author} {\bibfnamefont
  {N.}~\bibnamefont {Nagaosa}},\ and\ \bibinfo {author} {\bibfnamefont
  {Y.}~\bibnamefont {Tokura}},\ }\bibfield  {title} {\bibinfo {title}
  {Real-space observation of a two-dimensional skyrmion crystal},\ }\href
  {https://doi.org/10.1038/nature09124} {\bibfield  {journal} {\bibinfo
  {journal} {Nature}\ }\textbf {\bibinfo {volume} {465}},\ \bibinfo {pages}
  {901} (\bibinfo {year} {2010})}\BibitemShut {NoStop}%
\bibitem [{\citenamefont {Zhang}\ \emph {et~al.}(2016)\citenamefont {Zhang},
  \citenamefont {Zhou},\ and\ \citenamefont {Ezawa}}]{Zhang2016}%
  \BibitemOpen
  \bibfield  {author} {\bibinfo {author} {\bibfnamefont {X.}~\bibnamefont
  {Zhang}}, \bibinfo {author} {\bibfnamefont {Y.}~\bibnamefont {Zhou}},\ and\
  \bibinfo {author} {\bibfnamefont {M.}~\bibnamefont {Ezawa}},\ }\bibfield
  {title} {\bibinfo {title} {Antiferromagnetic skyrmion: Stability, creation
  and manipulation},\ }\href {https://doi.org/10.1038/srep24795} {\bibfield
  {journal} {\bibinfo  {journal} {Sci. Rep.}\ }\textbf {\bibinfo {volume}
  {6}},\ \bibinfo {pages} {24795} (\bibinfo {year} {2016})}\BibitemShut
  {NoStop}%
\bibitem [{\citenamefont {Liu}\ \emph {et~al.}(2018)\citenamefont {Liu},
  \citenamefont {Lake},\ and\ \citenamefont {Zang}}]{Liu:2018}%
  \BibitemOpen
  \bibfield  {author} {\bibinfo {author} {\bibfnamefont {Y.}~\bibnamefont
  {Liu}}, \bibinfo {author} {\bibfnamefont {R.~K.}\ \bibnamefont {Lake}},\ and\
  \bibinfo {author} {\bibfnamefont {J.}~\bibnamefont {Zang}},\ }\bibfield
  {title} {\bibinfo {title} {Binding a hopfion in a chiral magnet nanodisk},\
  }\href {https://doi.org/10.1103/PhysRevB.98.174437} {\bibfield  {journal}
  {\bibinfo  {journal} {Phys. Rev. B}\ }\textbf {\bibinfo {volume} {98}},\
  \bibinfo {pages} {174437} (\bibinfo {year} {2018})}\BibitemShut {NoStop}%
\bibitem [{\citenamefont {Sutcliffe}(2018)}]{Sutcliffe_2018}%
  \BibitemOpen
  \bibfield  {author} {\bibinfo {author} {\bibfnamefont {P.}~\bibnamefont
  {Sutcliffe}},\ }\bibfield  {title} {\bibinfo {title} {Hopfions in chiral
  magnets},\ }\href {https://doi.org/10.1088/1751-8121/aad521} {\bibfield
  {journal} {\bibinfo  {journal} {J. Phys. A Math.}\ }\textbf {\bibinfo
  {volume} {51}},\ \bibinfo {pages} {375401} (\bibinfo {year}
  {2018})}\BibitemShut {NoStop}%
\bibitem [{\citenamefont {Das}\ \emph {et~al.}(2019)\citenamefont {Das},
  \citenamefont {Tang}, \citenamefont {Hong}, \citenamefont {Gon{\c{c}}alves},
  \citenamefont {McCarter}, \citenamefont {Klewe}, \citenamefont {Nguyen},
  \citenamefont {G{\'o}mez-Ortiz}, \citenamefont {Shafer}, \citenamefont
  {Arenholz}, \citenamefont {Stoica}, \citenamefont {Hsu}, \citenamefont
  {Wang}, \citenamefont {Ophus}, \citenamefont {Liu}, \citenamefont {Nelson},
  \citenamefont {Saremi}, \citenamefont {Prasad}, \citenamefont {Mei},
  \citenamefont {Schlom}, \citenamefont {{\'I}{\~{n}}iguez}, \citenamefont
  {Garc{\'i}a-Fern{\'a}ndez}, \citenamefont {Muller}, \citenamefont {Chen},
  \citenamefont {Junquera}, \citenamefont {Martin},\ and\ \citenamefont
  {Ramesh}}]{Das2019}%
  \BibitemOpen
  \bibfield  {author} {\bibinfo {author} {\bibfnamefont {S.}~\bibnamefont
  {Das}}, \bibinfo {author} {\bibfnamefont {Y.~L.}\ \bibnamefont {Tang}},
  \bibinfo {author} {\bibfnamefont {Z.}~\bibnamefont {Hong}}, \bibinfo {author}
  {\bibfnamefont {M.~A.~P.}\ \bibnamefont {Gon{\c{c}}alves}}, \bibinfo {author}
  {\bibfnamefont {M.~R.}\ \bibnamefont {McCarter}}, \bibinfo {author}
  {\bibfnamefont {C.}~\bibnamefont {Klewe}}, \bibinfo {author} {\bibfnamefont
  {K.~X.}\ \bibnamefont {Nguyen}}, \bibinfo {author} {\bibfnamefont
  {F.}~\bibnamefont {G{\'o}mez-Ortiz}}, \bibinfo {author} {\bibfnamefont
  {P.}~\bibnamefont {Shafer}}, \bibinfo {author} {\bibfnamefont
  {E.}~\bibnamefont {Arenholz}}, \bibinfo {author} {\bibfnamefont {V.~A.}\
  \bibnamefont {Stoica}}, \bibinfo {author} {\bibfnamefont {S.-L.}\
  \bibnamefont {Hsu}}, \bibinfo {author} {\bibfnamefont {B.}~\bibnamefont
  {Wang}}, \bibinfo {author} {\bibfnamefont {C.}~\bibnamefont {Ophus}},
  \bibinfo {author} {\bibfnamefont {J.~F.}\ \bibnamefont {Liu}}, \bibinfo
  {author} {\bibfnamefont {C.~T.}\ \bibnamefont {Nelson}}, \bibinfo {author}
  {\bibfnamefont {S.}~\bibnamefont {Saremi}}, \bibinfo {author} {\bibfnamefont
  {B.}~\bibnamefont {Prasad}}, \bibinfo {author} {\bibfnamefont {A.~B.}\
  \bibnamefont {Mei}}, \bibinfo {author} {\bibfnamefont {D.~G.}\ \bibnamefont
  {Schlom}}, \bibinfo {author} {\bibfnamefont {J.}~\bibnamefont
  {{\'I}{\~{n}}iguez}}, \bibinfo {author} {\bibfnamefont {P.}~\bibnamefont
  {Garc{\'i}a-Fern{\'a}ndez}}, \bibinfo {author} {\bibfnamefont {D.~A.}\
  \bibnamefont {Muller}}, \bibinfo {author} {\bibfnamefont {L.~Q.}\
  \bibnamefont {Chen}}, \bibinfo {author} {\bibfnamefont {J.}~\bibnamefont
  {Junquera}}, \bibinfo {author} {\bibfnamefont {L.~W.}\ \bibnamefont
  {Martin}},\ and\ \bibinfo {author} {\bibfnamefont {R.}~\bibnamefont
  {Ramesh}},\ }\bibfield  {title} {\bibinfo {title} {Observation of
  room-temperature polar skyrmions},\ }\href
  {https://doi.org/10.1038/s41586-019-1092-8} {\bibfield  {journal} {\bibinfo
  {journal} {Nature}\ }\textbf {\bibinfo {volume} {568}},\ \bibinfo {pages}
  {368} (\bibinfo {year} {2019})}\BibitemShut {NoStop}%
\bibitem [{\citenamefont {Leonov}\ \emph {et~al.}(2014)\citenamefont {Leonov},
  \citenamefont {Dragunov}, \citenamefont {R\"o\ss{}ler},\ and\ \citenamefont
  {Bogdanov}}]{PhysRevE.90.042502}%
  \BibitemOpen
  \bibfield  {author} {\bibinfo {author} {\bibfnamefont {A.~O.}\ \bibnamefont
  {Leonov}}, \bibinfo {author} {\bibfnamefont {I.~E.}\ \bibnamefont
  {Dragunov}}, \bibinfo {author} {\bibfnamefont {U.~K.}\ \bibnamefont
  {R\"o\ss{}ler}},\ and\ \bibinfo {author} {\bibfnamefont {A.~N.}\ \bibnamefont
  {Bogdanov}},\ }\bibfield  {title} {\bibinfo {title} {Theory of skyrmion
  states in liquid crystals},\ }\href
  {https://doi.org/10.1103/PhysRevE.90.042502} {\bibfield  {journal} {\bibinfo
  {journal} {Phys. Rev. E}\ }\textbf {\bibinfo {volume} {90}},\ \bibinfo
  {pages} {042502} (\bibinfo {year} {2014})}\BibitemShut {NoStop}%
\bibitem [{\citenamefont {Duzgun}\ \emph
  {et~al.}(2018{\natexlab{b}})\citenamefont {Duzgun}, \citenamefont
  {Selinger},\ and\ \citenamefont {Saxena}}]{PhysRevE.97.062706}%
  \BibitemOpen
  \bibfield  {author} {\bibinfo {author} {\bibfnamefont {A.}~\bibnamefont
  {Duzgun}}, \bibinfo {author} {\bibfnamefont {J.~V.}\ \bibnamefont
  {Selinger}},\ and\ \bibinfo {author} {\bibfnamefont {A.}~\bibnamefont
  {Saxena}},\ }\bibfield  {title} {\bibinfo {title} {Comparing skyrmions and
  merons in chiral liquid crystals and magnets},\ }\href
  {https://doi.org/10.1103/PhysRevE.97.062706} {\bibfield  {journal} {\bibinfo
  {journal} {Phys. Rev. E}\ }\textbf {\bibinfo {volume} {97}},\ \bibinfo
  {pages} {062706} (\bibinfo {year} {2018}{\natexlab{b}})}\BibitemShut
  {NoStop}%
\bibitem [{\citenamefont {Foster}\ \emph {et~al.}(2019)\citenamefont {Foster},
  \citenamefont {Kind}, \citenamefont {Ackerman}, \citenamefont {Tai},
  \citenamefont {Dennis},\ and\ \citenamefont {Smalyukh}}]{Foster2019}%
  \BibitemOpen
  \bibfield  {author} {\bibinfo {author} {\bibfnamefont {D.}~\bibnamefont
  {Foster}}, \bibinfo {author} {\bibfnamefont {C.}~\bibnamefont {Kind}},
  \bibinfo {author} {\bibfnamefont {P.~J.}\ \bibnamefont {Ackerman}}, \bibinfo
  {author} {\bibfnamefont {J.~S.~B.}\ \bibnamefont {Tai}}, \bibinfo {author}
  {\bibfnamefont {M.~R.}\ \bibnamefont {Dennis}},\ and\ \bibinfo {author}
  {\bibfnamefont {I.~I.}\ \bibnamefont {Smalyukh}},\ }\bibfield  {title}
  {\bibinfo {title} {Two-dimensional skyrmion bags in liquid crystals and
  ferromagnets},\ }\href {https://www.nature.com/articles/s41567-019-0476-x}
  {\bibfield  {journal} {\bibinfo  {journal} {Nat. Phys.}\ }\textbf {\bibinfo
  {volume} {15}} (\bibinfo {year} {2019})}\BibitemShut {NoStop}%
\bibitem [{\citenamefont {Teixeira}\ \emph {et~al.}(2021)\citenamefont
  {Teixeira}, \citenamefont {Castillo-Sepúlveda}, \citenamefont {Rizzi},
  \citenamefont {Nunez}, \citenamefont {Troncoso}, \citenamefont {Altbir},
  \citenamefont {Fonseca},\ and\ \citenamefont
  {Carvalho-Santos}}]{Teixeira_2021}%
  \BibitemOpen
  \bibfield  {author} {\bibinfo {author} {\bibfnamefont {A.~W.}\ \bibnamefont
  {Teixeira}}, \bibinfo {author} {\bibfnamefont {S.}~\bibnamefont
  {Castillo-Sepúlveda}}, \bibinfo {author} {\bibfnamefont {L.~G.}\
  \bibnamefont {Rizzi}}, \bibinfo {author} {\bibfnamefont {A.~S.}\ \bibnamefont
  {Nunez}}, \bibinfo {author} {\bibfnamefont {R.~E.}\ \bibnamefont {Troncoso}},
  \bibinfo {author} {\bibfnamefont {D.}~\bibnamefont {Altbir}}, \bibinfo
  {author} {\bibfnamefont {J.~M.}\ \bibnamefont {Fonseca}},\ and\ \bibinfo
  {author} {\bibfnamefont {V.~L.}\ \bibnamefont {Carvalho-Santos}},\ }\bibfield
   {title} {\bibinfo {title} {Motion-induced inertial effects and topological
  phase transitions in skyrmion transport},\ }\href
  {https://doi.org/10.1088/1361-648X/abfb8c} {\bibfield  {journal} {\bibinfo
  {journal} {J. Phys. Cond. Matt.}\ }\textbf {\bibinfo {volume} {33}},\
  \bibinfo {pages} {265403} (\bibinfo {year} {2021})}\BibitemShut {NoStop}%
\bibitem [{\citenamefont {Krause}\ and\ \citenamefont
  {Wiesendanger}(2016)}]{Krause2016}%
  \BibitemOpen
  \bibfield  {author} {\bibinfo {author} {\bibfnamefont {S.}~\bibnamefont
  {Krause}}\ and\ \bibinfo {author} {\bibfnamefont {R.}~\bibnamefont
  {Wiesendanger}},\ }\bibfield  {title} {\bibinfo {title} {Skyrmionics gets
  hot},\ }\href {https://doi.org/10.1038/nmat4615} {\bibfield  {journal}
  {\bibinfo  {journal} {Nat. Mater.}\ }\textbf {\bibinfo {volume} {15}},\
  \bibinfo {pages} {493} (\bibinfo {year} {2016})}\BibitemShut {NoStop}%
\bibitem [{\citenamefont {Shen}\ and\ \citenamefont
  {Dierking}(2020)}]{Shen2020}%
  \BibitemOpen
  \bibfield  {author} {\bibinfo {author} {\bibfnamefont {Y.}~\bibnamefont
  {Shen}}\ and\ \bibinfo {author} {\bibfnamefont {I.}~\bibnamefont
  {Dierking}},\ }\bibfield  {title} {\bibinfo {title} {Dynamic dissipative
  solitons in nematics with positive anisotropies},\ }\href
  {https://pubs.rsc.org/en/content/articlelanding/2020/sm/d0sm00676a}
  {\bibfield  {journal} {\bibinfo  {journal} {Soft Matter}\ }\textbf {\bibinfo
  {volume} {16}} (\bibinfo {year} {2020})}\BibitemShut {NoStop}%
\bibitem [{\citenamefont {Wu}\ and\ \citenamefont {Smalyukh}(2022)}]{Wu2022}%
  \BibitemOpen
  \bibfield  {author} {\bibinfo {author} {\bibfnamefont {J.-S.}\ \bibnamefont
  {Wu}}\ and\ \bibinfo {author} {\bibfnamefont {I.~I.}\ \bibnamefont
  {Smalyukh}},\ }\bibfield  {title} {\bibinfo {title} {Hopfions, heliknotons,
  skyrmions, torons and both abelian and nonabelian vortices in chiral liquid
  crystals},\ }\href {https://doi.org/10.1080/21680396.2022.2040058} {\bibfield
   {journal} {\bibinfo  {journal} {Liq. Cryst. Rev.}\ ,\ \bibinfo {pages} {1}}
  (\bibinfo {year} {2022})}\BibitemShut {NoStop}%
\bibitem [{\citenamefont {Sohn}\ and\ \citenamefont
  {Smalyukh}(2020)}]{Sohn2020}%
  \BibitemOpen
  \bibfield  {author} {\bibinfo {author} {\bibfnamefont {H.~R.~O.}\
  \bibnamefont {Sohn}}\ and\ \bibinfo {author} {\bibfnamefont {I.~I.}\
  \bibnamefont {Smalyukh}},\ }\bibfield  {title} {\bibinfo {title}
  {Electrically powered motions of toron crystallites in chiral liquid
  crystals},\ }\href {https://doi.org/10.1073/pnas.1922198117} {\bibfield
  {journal} {\bibinfo  {journal} {Proc. Natl. Acad. Sci. USA}\ }\textbf
  {\bibinfo {volume} {117}},\ \bibinfo {pages} {6437} (\bibinfo {year}
  {2020})},\ \Eprint
  {https://arxiv.org/abs/https://www.pnas.org/doi/pdf/10.1073/pnas.1922198117}
  {https://www.pnas.org/doi/pdf/10.1073/pnas.1922198117} \BibitemShut {NoStop}%
\bibitem [{\citenamefont {Li}\ \emph {et~al.}(2020)\citenamefont {Li},
  \citenamefont {Shen}, \citenamefont {Bai}, \citenamefont {Wang},
  \citenamefont {Zhang}, \citenamefont {Xia}, \citenamefont {Ezawa},
  \citenamefont {Tretiakov}, \citenamefont {Xu}, \citenamefont {Mruczkiewicz},
  \citenamefont {Krawczyk}, \citenamefont {Xu}, \citenamefont {Evans},
  \citenamefont {Chantrell},\ and\ \citenamefont {Zhou}}]{Li2020}%
  \BibitemOpen
  \bibfield  {author} {\bibinfo {author} {\bibfnamefont {X.}~\bibnamefont
  {Li}}, \bibinfo {author} {\bibfnamefont {L.}~\bibnamefont {Shen}}, \bibinfo
  {author} {\bibfnamefont {Y.}~\bibnamefont {Bai}}, \bibinfo {author}
  {\bibfnamefont {J.}~\bibnamefont {Wang}}, \bibinfo {author} {\bibfnamefont
  {X.}~\bibnamefont {Zhang}}, \bibinfo {author} {\bibfnamefont
  {J.}~\bibnamefont {Xia}}, \bibinfo {author} {\bibfnamefont {M.}~\bibnamefont
  {Ezawa}}, \bibinfo {author} {\bibfnamefont {O.~A.}\ \bibnamefont
  {Tretiakov}}, \bibinfo {author} {\bibfnamefont {X.}~\bibnamefont {Xu}},
  \bibinfo {author} {\bibfnamefont {M.}~\bibnamefont {Mruczkiewicz}}, \bibinfo
  {author} {\bibfnamefont {M.}~\bibnamefont {Krawczyk}}, \bibinfo {author}
  {\bibfnamefont {Y.}~\bibnamefont {Xu}}, \bibinfo {author} {\bibfnamefont
  {R.~F.~L.}\ \bibnamefont {Evans}}, \bibinfo {author} {\bibfnamefont {R.~W.}\
  \bibnamefont {Chantrell}},\ and\ \bibinfo {author} {\bibfnamefont
  {Y.}~\bibnamefont {Zhou}},\ }\bibfield  {title} {\bibinfo {title} {Bimeron
  clusters in chiral antiferromagnets},\ }\href
  {https://doi.org/10.1038/s41524-020-00435-y} {\bibfield  {journal} {\bibinfo
  {journal} {Npj Comput. Mater.}\ }\textbf {\bibinfo {volume} {6}},\ \bibinfo
  {pages} {169} (\bibinfo {year} {2020})}\BibitemShut {NoStop}%
\bibitem [{\citenamefont {Plimpton}(1995)}]{PLIMPTON19951}%
  \BibitemOpen
  \bibfield  {author} {\bibinfo {author} {\bibfnamefont {S.}~\bibnamefont
  {Plimpton}},\ }\bibfield  {title} {\bibinfo {title} {Fast parallel algorithms
  for short-range molecular dynamics},\ }\href
  {https://www.sciencedirect.com/science/article/pii/S002199918571039X}
  {\bibfield  {journal} {\bibinfo  {journal} {J. Comp. Phys.}\ }\textbf
  {\bibinfo {volume} {117}},\ \bibinfo {pages} {1} (\bibinfo {year}
  {1995})}\BibitemShut {NoStop}%
\bibitem [{\citenamefont {Verlet}(1967)}]{PhysRev.159.98}%
  \BibitemOpen
  \bibfield  {author} {\bibinfo {author} {\bibfnamefont {L.}~\bibnamefont
  {Verlet}},\ }\bibfield  {title} {\bibinfo {title} {Computer "experiments" on
  classical fluids. i. thermodynamical properties of lennard-jones molecules},\
  }\href {https://doi.org/10.1103/PhysRev.159.98} {\bibfield  {journal}
  {\bibinfo  {journal} {Phys. Rev.}\ }\textbf {\bibinfo {volume} {159}},\
  \bibinfo {pages} {98} (\bibinfo {year} {1967})}\BibitemShut {NoStop}%
\bibitem [{\citenamefont {Bechinger}\ \emph {et~al.}(2016)\citenamefont
  {Bechinger}, \citenamefont {Di~Leonardo}, \citenamefont {L\"owen},
  \citenamefont {Reichhardt}, \citenamefont {Volpe},\ and\ \citenamefont
  {Volpe}}]{RevModPhys.88.045006}%
  \BibitemOpen
  \bibfield  {author} {\bibinfo {author} {\bibfnamefont {C.}~\bibnamefont
  {Bechinger}}, \bibinfo {author} {\bibfnamefont {R.}~\bibnamefont
  {Di~Leonardo}}, \bibinfo {author} {\bibfnamefont {H.}~\bibnamefont
  {L\"owen}}, \bibinfo {author} {\bibfnamefont {C.}~\bibnamefont {Reichhardt}},
  \bibinfo {author} {\bibfnamefont {G.}~\bibnamefont {Volpe}},\ and\ \bibinfo
  {author} {\bibfnamefont {G.}~\bibnamefont {Volpe}},\ }\bibfield  {title}
  {\bibinfo {title} {Active particles in complex and crowded environments},\
  }\href {https://doi.org/10.1103/RevModPhys.88.045006} {\bibfield  {journal}
  {\bibinfo  {journal} {Rev. Mod. Phys.}\ }\textbf {\bibinfo {volume} {88}},\
  \bibinfo {pages} {045006} (\bibinfo {year} {2016})}\BibitemShut {NoStop}%
\end{thebibliography}%
\vspace{0.5cm}

\end{document}